%% file: manuscript.tex
\documentclass[aps,twocolumn,pra,superscriptaddress]{revtex4-2}

\usepackage{graphics}
\usepackage{graphicx}
\usepackage{bm}
\usepackage{latexsym}
\usepackage{color, soul}
\usepackage{mathrsfs}
\usepackage{braket}
\usepackage{enumerate}
\usepackage[usenames,dvipsnames]{xcolor}
\usepackage[colorlinks=true,citecolor=blue,linkcolor=blue,urlcolor=blue,backref=true]{hyperref}
\usepackage{float}
\usepackage[raggedright,tight]{subfigure}  
\usepackage{amsmath}
\usepackage{amsfonts}
\usepackage{braket}
\usepackage[normalem]{ulem}
\usepackage{cancel}
\usepackage{mathtools}
\usepackage{appendix}
\usepackage{comment}
\usepackage{bm}

\usepackage{bm}

\def\r{{\bm r}}
\def\R{{\bm R}}
\def\Ro{{\bm \rho}}

\def\GG{{\bm G}}
\def\Gb{{\bm G_b}}
\def\alpb{{{{\bm{\alpha}}}}}
\def\kp{{\bm k_{\parallel}}}

\begin{document}

\title{Long-range quantum emitter interactions mediated by a non-local metasurface: Application to qubit-qubit entanglement}

\author{Hannah Riley}
  \affiliation{Department of Physics and Astronomy, The University of Manchester, Oxford Road, Manchester, M13 9PL, United Kingdom}
  \affiliation{Institute of Materials Research and Engineering (IMRE), Agency for Science, Technology and Research (A*STAR), 2 Fusionopolis Way, Innovis \#08-03, Singapore 138634, Republic of Singapore}
\author{Emmanuel Lassalle}
\email{emmanuel\_lassalle@a-star.edu.sg}
  \affiliation{Institute of Materials Research and Engineering (IMRE), Agency for Science, Technology and Research (A*STAR), 2 Fusionopolis Way, Innovis \#08-03, Singapore 138634, Republic of Singapore}
\author{Diego Romero Abujetas}
    \affiliation{Department of Physics, University of Fribourg, 1700 Fribourg, Switzerland}
    \affiliation{Department of Mathematics, Universidad de Castilla-La Mancha, Avenida Carlos III s/n, 45071 Toledo, Spain}
\author{Adam Stokes}
    \affiliation{School of Mathematics, Statistics, and Physics, Newcastle University, Newcastle upon Tyne NE1 7RU, United Kingdom}
\author{Ramon Paniagua-Dominguez}
\email{ramon.paniagua@csic.es}
    \affiliation{Institute of Materials Research and Engineering (IMRE), Agency for Science, Technology and Research (A*STAR), 2 Fusionopolis Way, Innovis \#08-03, Singapore 138634, Republic of Singapore}
\author{Ahsan Nazir}\email{ahsan.nazir@manchester.ac.uk}
  \affiliation{Department of Physics and Astronomy, The University of Manchester, Oxford Road, Manchester, M13 9PL, United Kingdom}
\date{\today}

\begin{abstract}
Scalable quantum technologies demand long-range interactions between many distant quantum emitters (QEs). We introduce non-local metasurfaces supporting bound-states-in-the-continuum (BICs) as a promising platform to achieve this goal. 
We show that efficient QE interactions depend almost entirely on emitter–BIC coupling efficiencies ($\beta$-factors), which in our system can exceed 80\% even without additional mode engineering.
These values rival those of 1D waveguides but are achieved here in a geometry that naturally accommodates large 2D QE arrays.
Using this platform, we explore entanglement generation between two remote QEs, finding that it develops faster than in free space, is significantly amplified, and persists over separations spanning several emission wavelengths.
Optimal inter-QE interactions require large $\beta$-factors but only moderately small Purcell factors, both 
within experimentally achievable ranges. Our results establish non-local metasurfaces as a practical and scalable platform for leading-edge 
quantum nanophotonics.
\end{abstract}

\maketitle

\section{Introduction}

Long-range interactions in quantum systems are highly sought after because they enable collective quantum phenomena and play a central role in advancing emerging quantum technologies \cite{defenu2023long}. In particular, they facilitate robust and scalable entanglement between spatially separated quantum emitters (QEs), a key requirement 
for quantum computing \cite{laucht2021roadmap}. Such interactions also circumvent the need to fabricate sub-wavelength QE arrays \cite{shah2024quantum} --- a process that is experimentally demanding, and undesirable when individual optical, electrical, or magnetic control over each QE is required, since dense packing inevitably leads to cross-talk.

Photon-mediated interactions can be enhanced by tailoring structured photonic environments \cite{Lodahl2015}, which has inspired extensive theoretical \cite{hughes2005modified, hou2014dissipation, biehs2017qubit, miguel2022inverse} and experimental \cite{samutpraphoot2020strong, trebbia2022tailoring, Holland2023} studies. In particular, one-dimensional (1D) waveguides \cite{sheremet2023waveguide} have been proposed to enable long-range QE–QE coupling and entanglement creation \cite{dzsotjan2010quantum, martin2011dissipation, gonzalez2011entanglement, zheng2013persistent, shahmoon2013nonradiative, gonzalez2013mesoscopic}, with a recent experiment demonstrating enhanced coupling of quantum dots via a photonic crystal waveguide \cite{Tiranov2023}. However, the intrinsic 1D geometry of such platforms poses a challenge to scalability, which motivates the exploration of two-dimensional (2D) architectures capable of supporting entanglement across planar QE arrays.

Metasurfaces, comprising planar arrays of engineered nanoparticles (“meta-atoms”), enable sub-wavelength control of light into the quantum regime \cite{Ji2023,Zheng2023}, while also offering scalability, along with advantages in fabrication and integration \cite{Kildishev2013,Yu2014,Soukoulis2011,GomezRivas2019,metasurfacemainstream2023}. A theoretical study by Jha et al.~\cite{Jha2017} demonstrated the potential of metasurfaces to effectively facilitate entanglement generation by considering atomic qubits in the metasurface \emph{far-field}. However, suspending the QEs in the far-field requires additional systems, such as optical tweezers \cite{bloch2023trapping}, 
and the metasurface design was limited to two atoms at specific positions, making it unsuitable for QE arrays. More recently, Zundel et al.~\cite{Zundel2022} studied long-range interactions 
mediated by lattice resonances in the \emph{near-field} of a metallic metasurface, which are compatible with planar QE arrays. 
While near-field coupling to such modes offers a promising architecture for compact integration \cite{Aharonovich2022,do2024room,an2025dielectric}, it also introduces losses inherent to metallic systems.

\begin{figure*}
    \centering
        \centering
\hspace*{-1cm}\includegraphics[width=0.9\linewidth]{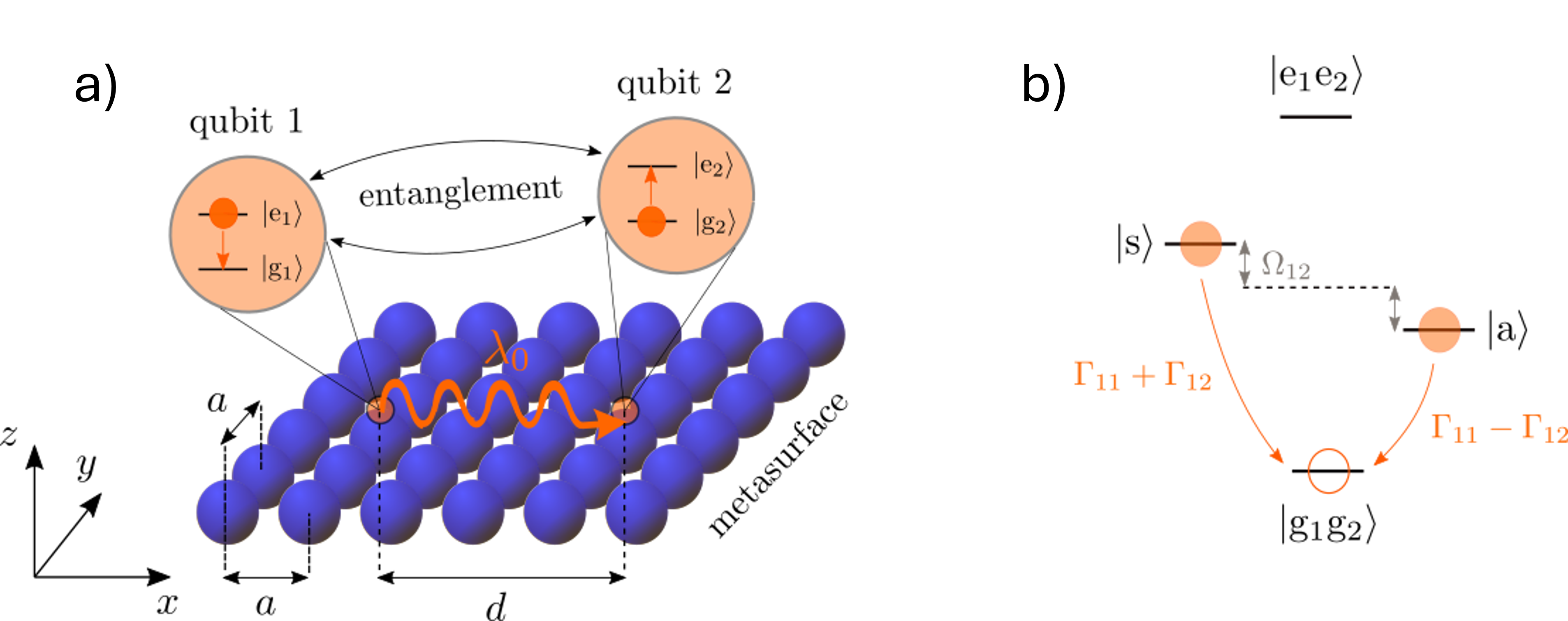} 
        \caption{
        (a) Schematic illustrating long-range metasurface-mediated interactions between two qubits. Initially uncorrelated QEs (qubits) spontaneously emit into delocalized BIC modes, resulting in entanglement over distances much larger than the wavelength of emitted photons. (b) Dicke states of the two emitter system are connected by decay channels with rates $\Gamma_{11}\pm \Gamma_{12}$. Initially QE1 is excited, such that the two-QE system is in an equal superposition of symmetric state $\ket{\mathrm{s}}$ and the antisymmetric state $\ket{\mathrm{a}}$.}
        \label{fig:schematics}
\end{figure*}

Here, we propose instead to use \emph{non-local} metasurfaces that support {\em bound-states-in-the-continuum} (BICs).
BICs are spatially delocalized wave states spectrally embedded in the radiation continuum but decoupled from it, preventing energy leakage and yielding infinite lifetimes and quality factors ($Q$-factors) \cite{Hsu2016,Muljarov2021,Koshelev2023, huang2023resonant, kang2023applications}.
Although they cannot be excited via the far field, BICs can be efficiently accessed using localized near-field sources \cite{kodigala2017lasing, ha2018directional, wu2020room, wu2021bound}.
In practice, finite and imperfectly fabricated metasurfaces host quasi-BICs with very high yet finite $Q$-factors. 

We find that non-local metasurfaces supporting BICs can mediate long-range interactions and entanglement between QEs over separations exceeding several emission wavelengths, persisting indefinitely in the ideal case of an infinite metasurface with perfect spectral and spatial matching between the QEs and the BIC mode. QE-BIC coupling efficiencies, quantified by $\beta$-factors \cite{lecamp2007very,manga2007single,arcari2014near}, are found to be exceptionally high for a 2D platform, exceeding 80\% and potentially improvable through further metasurface optimization. For configurations that maximise QE-BIC coupling, the resulting QE-QE entanglement depends almost entirely on the $\beta$-factor, increasing monotonically with it. This is counterintuitive from a classical perspective, where one would expect that the source dipole field would spread across the metasurface plane such that only a small fraction reaches the target emitter, significantly reducing their coupling and  entanglement. Moreover, the metasurface induces a notable Purcell effect that accelerates entanglement generation, a key advantage for fast quantum operations.


The article is organized as follows. We first introduce the composite QE-metasurface system, where the QEs are described using a Lindblad master equation. The metasurface-modified single and collective QE decay rates are expressed in terms of the local and cross densities of states (LDOS and CDOS) \cite{canaguier2016quantum}. We compute these quantities for two distinct BICs --- electric dipole (ED-BIC) and magnetic dipole (MD-BIC) --- which exhibit different optimal near-field coupling conditions. A combined analytical and numerical analysis yields $\beta$-factors that quantify QE–BIC coupling and determine the resulting dynamics. To quantify QE–QE entanglement generated through spontaneous emission \cite{tanas2004entangling}, we employ concurrence and derive an approximate expression in terms of $\beta$- and Purcell factors, which can serve as a design guide for metasurfaces interfaced with QE arrays.

\section{Results}\label{section:Results}

\subsection{The QE-metasurface system}

We consider a pair of two-level QEs, labeled $\mu= 1, 2$ (Fig.~\ref{fig:schematics}a), with ground and excited states $\ket{\mathrm{g}_\mu}$ and $\ket{\mathrm{e}_\mu}$, respectively, and identical transition energy $\hbar \omega_0$, where $\hbar$ is the reduced Planck constant. The corresponding transition wavelength is $\lambda_0 =2\pi c/\omega_0$, where $c$ is the speed of light. We assume identical transition dipole moment magnitudes, $p_1=p_2 =p$, but allow different orientations; ${\hat {\bf p}}_1 \neq {\hat {\bf p}}_2$ in general. Initially, QE1 is prepared in its excited state ($\ket{\mathrm{e}_1}$), while QE2 is in its ground state ($\ket{\mathrm{g}_2}$).

\begin{figure*}
    \centering
        \centering
\includegraphics[width=0.9\linewidth]{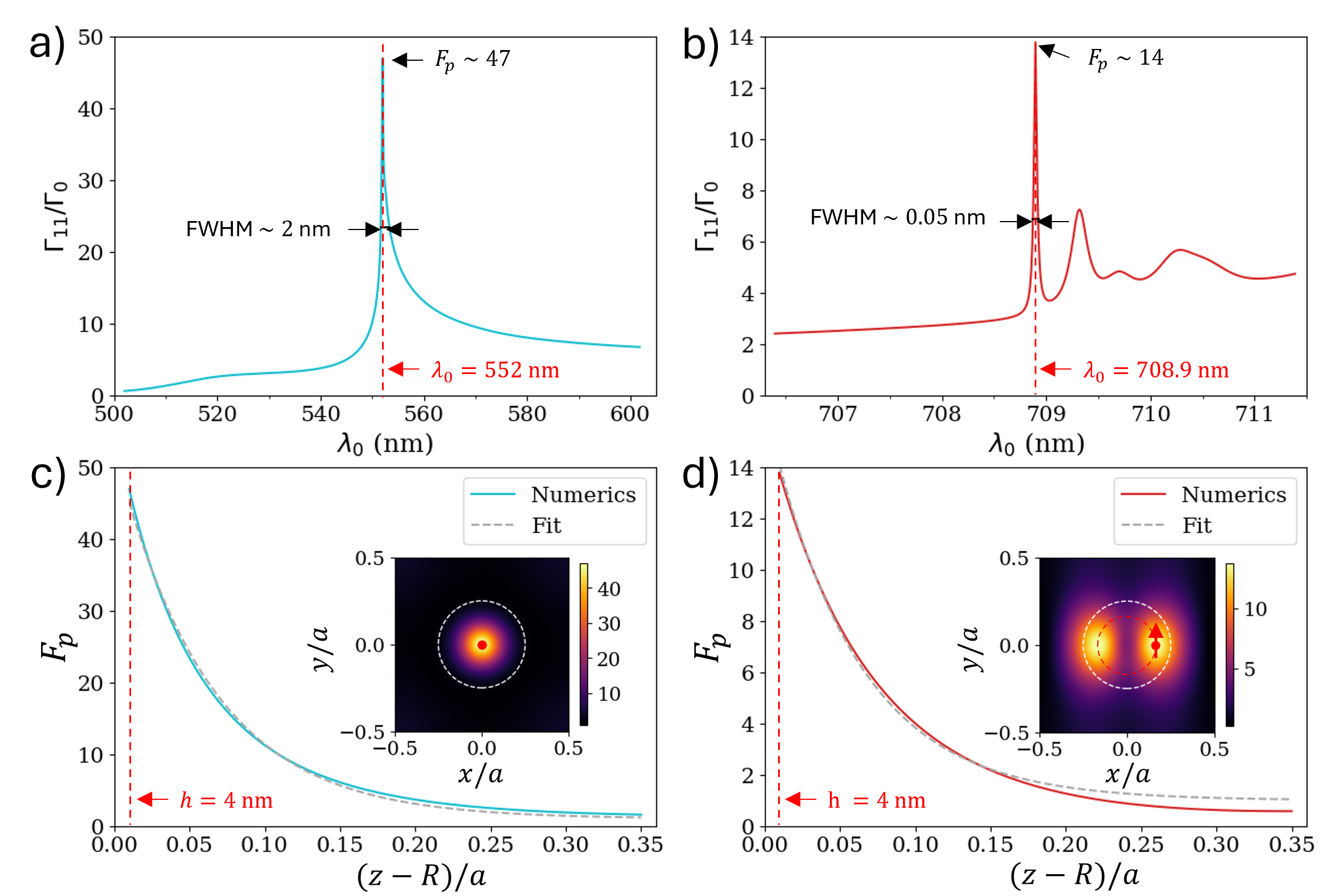} 
        \caption{
        Normalized single-emitter decay rates $\Gamma_{11}/\Gamma_0$ (LDOS) as a function of emitter wavelength $\lambda_0$ for the ED-BIC (a) and MD-BIC (b) at emitter height $h=z-R=4\,$nm above the metasurface, and Purcell factor $F_p$ as a function of $z$ for the ED-BIC at $\lambda_\text{BIC}=552.0\,$nm (c) and for MD-BIC at $\lambda_\text{BIC}=708.9\,$nm (d). In (a) and (b), the emitter's position and orientation are chosen to maximise the coupling (see main text). Dashed grey lines are exponential fits of the form $1+A\mathrm{e}^{-B(z-R)/a}$ with fitting parameters $A=51.80$ and $B=16.05$ (c), and $A=15.47$ and $B=16.97$ (d).
        Insets in (c) and (d): Purcell factors as a function of lateral position in the $xy$-plane at $h=4\,$nm for an emitter with orientation along the $z$-axis (out-of-plane) (c) and orientation along the $y$-axis (in-plane) (d). Red dots and arrows show the lateral positions and orientations of maximum coupling considered in (a) and (b). Dashed white circles show a nanosphere cross-section. The dashed red circle in (d) shows the positions where maximum coupling occurs.
        } 
        \label{fig:EDBIC_LDOS_vs_wl}
\end{figure*}

The metasurface under study is depicted in Fig.~\ref{fig:schematics}(a), comprising a square array of $21\times 21$ silicon nanospheres in the $xy$-plane, with lattice constant $a = 400\,$nm, radius $R=100\,$nm, and refractive index $n=3.5$~\cite{Abujetas21}. 
This finite size balances computational tractability and physical realism. 
The QEs are positioned at a vertical height $h=z-R=4\,$nm above the metasurface, enabling efficient near-field coupling to its optical resonances. 

The two-QE system can be described in terms of four (Dicke) levels, 
with two decay channels of rates $\Gamma_{11}\pm \Gamma_{12}$, separated by energy $\Omega_{12}$ \cite{tanas2004entangling, gonzalez2011entanglement, Jha2017} (shown in Fig.~\ref{fig:schematics}(b)). The coefficients $\Gamma_{\mu\nu}$ and $\Omega_{\mu\nu}$ are expressed in terms of the classical Green's tensor  $\bf{G}$ of Maxwell's equations, evaluated at the QE positions ${\bf R}_\mu$ and ${\bf R}_\nu$, as \cite{dung2002resonant}
\begin{equation}
\Gamma_{\mu \nu} 
= \frac{2  p_i p_j  \omega_0^2}{ \hbar \varepsilon_0 c^2}  \Im(G_{ij}(\bf{R}_{\mu},\bf{R}_{\nu},\omega_0)),
\label{eq:gamma_coll}
\end{equation}
and 
\begin{equation}
\Omega_{\mu\neq\nu} = - \frac{\omega_0^2}{\hbar \varepsilon_0 c^2} p_i p_j \Re(G_{ij}(\bf{R}_{\mu},\bf{R}_{\nu},\omega_0)),
\label{eqn:collective_coupling_rate}
\end{equation}
where $\varepsilon_0$ is the vacuum permittivity and repeated Cartesian indices $i, j$ are summed (Einstein convention). 
We denote $\Gamma_{11}$ ($\Gamma_{22}$) as the \emph{single-emitter decay rate} of QE 1 (QE 2), $\Gamma_{12}$ as the \emph{collective decay rate}, and $\Omega_{12}$ as the \emph{collective coupling rate}. 
In the dissipative regime, ($\Omega_{12}\simeq 0$, justified later), the effective coupling between the two QEs is quantified by $\Gamma_{12}$. More generally, when normalized by the free space dipole decay rate $\Gamma_0=\omega_0^3p^2/(3\pi\varepsilon_0\hbar c^3)$, $\Gamma_{\mu\mu}$ and $\Gamma_{12}$ depend only on the 
electromagnetic environment. They are related respectively to the LDOS and CDOS, which are proportional to the imaginary part of ${\bf G}$. Unlike the LDOS, the CDOS can take negative values, though the inequality $|\Gamma_{12}| \leq \sqrt{\Gamma_{11}}\sqrt{\Gamma_{22}}$, stemming from properties of the Green's tensor, is always satisfied \cite{canaguier2016quantum}. 

In free space, significant dipole-dipole interactions only occur for separations $d < \lambda_0$ \cite{stokes2018master}, 
since 
radiation from one dipole spreads isotropically and only a small fraction reaches the target.
By contrast, we will show that QE–metasurface platforms can be engineered to enhance emission through resonant modes, enabling dipole-dipole interactions over macroscopic separations $d>\lambda_0$. As mentioned earlier, we focus here on two types of optical resonance: the electric-dipole BIC (ED-BIC) and the magnetic-dipole BIC (MD-BIC)C~\cite{Abujetas21}.
In each case, the LDOS and the CDOS are computed numerically using the Python package SMUTHI (see Supplemental Material (SM), Sec.~I \cite{SM}). 

To model the two-QE system dynamics we use the reduced density matrix $\rho(t)$, which satisfies a Lindblad master equation 
of the same form as in free space~\cite{Agarwal2012}, 
but with coefficients given by Eqs.~(\ref{eq:gamma_coll}) and (\ref{eqn:collective_coupling_rate}) to account for the metasurface environment (details are provided in SM, Sec.~II \cite{SM}). This equation is valid in the weak-coupling regime. We have verified that at $h=4\,$nm the QEs remain weakly-coupled to the metasurface provided dipole moment magnitudes do not exceed $\sim90\,$Debye (see SM, Sec.~III \cite{SM}). Consequently, in solving the master equation with the Python package QuTIP, we restrict to $p\sim3\,$Debye (see SM, Sec.~I \cite{SM}). 

\begin{figure*}
     \centering
         \includegraphics[width=0.74\linewidth]{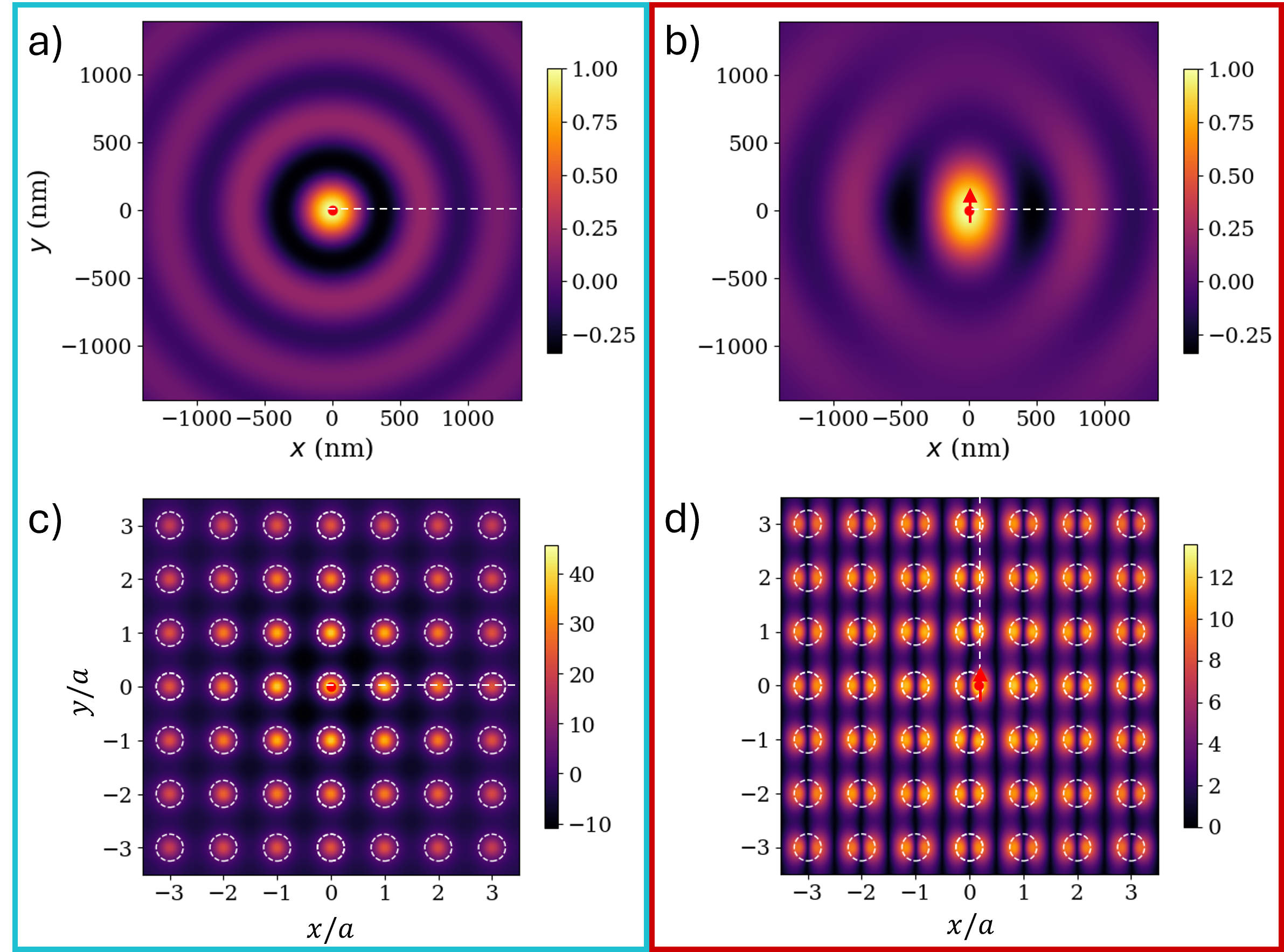} 
 \caption{
(a,b) Normalized collective decay rate $\Gamma_{12}/\Gamma_0$ (CDOS) in free space, with the first emitter fixed at $(x=0, y=0)$ (red dot) and the second emitter scanned laterally across the $xy$-plane. In (a)
both dipoles are oriented along $z$, in (b) they are oriented along $y$. 
(c) $\Gamma_{12}/\Gamma_0$ in the presence of the metasurface for the ED-BIC, with the first emitter again at $(0,0)$. 
(d) Collective decay rate for the MD-BIC, plotted as the absolute value $|\Gamma_{12}|/\Gamma_0$ to enhance visual contrast (the full signed distribution is given in SM, Sec.~V \cite{SM}). Here the first emitter is located at $(x=0.163a, y=0)$. 
In the left (resp. right) column both emitters are oriented along the $z$-axis (out-of-plane) (resp. $y$-axis (in-plane)), with transition wavelength $\lambda_0 = 552.0\,$nm (resp. $\lambda_0 = 708.9\,$nm). 
In the metasurface cases (c,d), both emitters are positioned at a height $h = 4\,$nm above the array. 
The spatial regions shown for free space [(a,b)] match those for the metasurface [(c,d)], covering $7 \times 7$ nanospheres centered on the central element, for direct comparison. 
Dashed white circles indicate nanosphere cross-sections. 
Dashed white lines mark the cuts used in later analysis: along $x$ at $y=0$ in (a-c) and along $y$ at $x=0.163a$ in (d).}        
         \label{fig:heat-maps}
 \end{figure*}

\subsection{Metasurface local density of states (LDOS)}

Figs.~\ref{fig:EDBIC_LDOS_vs_wl}(a) and (b) show the normalized single-emitter decay rates $\Gamma_{11}/\Gamma_0$, also referred to as the LDOS, as a function of $\lambda_0$ near the resonance wavelengths of the ED-BIC and MD-BIC, respectively. The maximum value, achieved when $\lambda_0$ matches the BIC resonance, corresponds to the Purcell factor $F_p$. This is plotted in Figs.~\ref{fig:EDBIC_LDOS_vs_wl}(c) (ED-BIC) and (d) (MD-BIC) as a function of the QE vertical position above the metasurface. At $h=4\,$nm, $F_p=46.9$ for the ED-BIC, occurring at resonance wavelength $\lambda_0=552\,$nm, and $F_p=13.7$ for the MD-BIC, occurring at $\lambda_0=708.9\,$nm. Although the MD-BIC possesses a smaller Purcell factor, it exhibits a much narrower linewidth (smaller full width at half maximum (FWHM)), corresponding to a larger $Q$-factor, which becomes relevant in the strong-coupling regime 
(see SM, Sec.~III \cite{SM}). 
Insets in Figs.~\ref{fig:EDBIC_LDOS_vs_wl}(c) and (d) show the lateral dependence of $F_p$ at fixed emitter orientations.

The emitter orientation and lateral position are chosen to maximize coupling. For the ED-BIC, the emitter must be placed at the center of a nanosphere with an out-of-plane orientation ($z$-direction), as illustrated in  Fig.~\ref{fig:EDBIC_LDOS_vs_wl}(c)). The MD-BIC allows greater flexibility: the emitter can be positioned anywhere on a circle of radius $r=0.163a$ centered on a nanosphere, with an in-plane orientation tangential to this circle (Fig.~\ref{fig:EDBIC_LDOS_vs_wl}(d)). Note that small positional changes will not affect the coupling drastically; for example, the Purcell factor decreases by less than 10\% for an emitter within a circle of diameter $30\,$nm around the optimum position for the ED-BIC ($40\,$nm for the MD-BIC). Such positioning accuracy may be well within reach experimentally \cite{li2021scalable, pambudi2024deterministic}.

For both BICs, the dependence of QE-BIC coupling (Purcell factor) on vertical position $z$ is essentially exponential and well described by
the function $1+A\mathrm{e}^{-B(z-R)/a}$, where $A$ and $B$ are dimensionless fitting parameters [dashed grey lines Figs.~\ref{fig:EDBIC_LDOS_vs_wl}(c) and (d)]. This behaviour is characteristic of near-field coupling to a surface-bound mode
and indicates that varying $z$ provides predictable control over the coupling strength.
It also allows prediction of the variation of the coupling with vertical position changes. Contrary to the lateral position, the Purcell factor is much more sensitive to variation in height: it decreases by 10\% when the height is increased by $\sim 2\,$nm for the ED-BIC ($\sim 3\,$nm for the MD-BIC) from the value $h=4\,$nm considered here. However, the height can be controlled very precisely with certain materials, such as 2D materials \cite{do2024room}.

\subsection{Metasurface cross density of states (CDOS)}

\begin{figure*}
       \centering
        \includegraphics[width=0.8\linewidth]{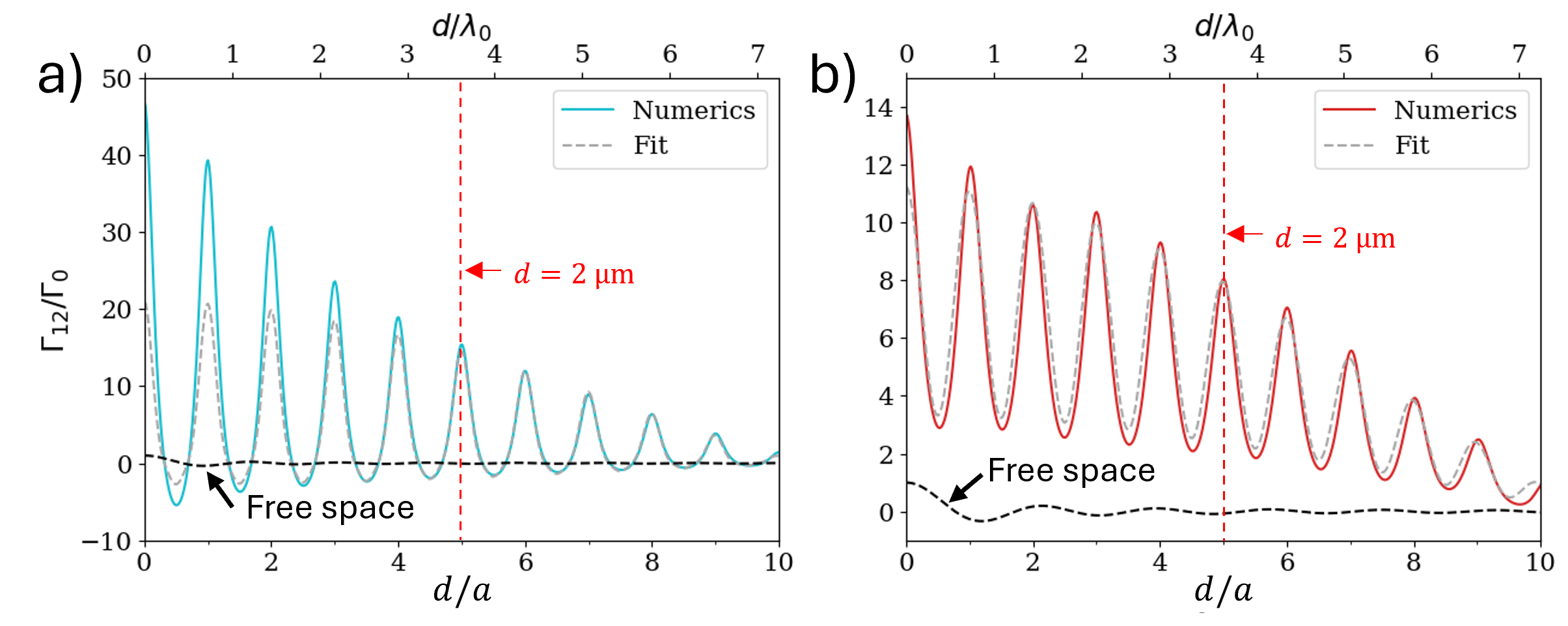} 
        \caption{
        Normalized collective decay rate $\Gamma_{12}/\Gamma_0$ (CDOS) as a function of emitter separation $d$ for the ED-BIC (a) 
        and for the MD-BIC (b), for cross-sections indicated by the white dashed lines in Fig.~\ref{fig:heat-maps}(c) and (d), respectively. 
        The dashed grey curves are fits to the numerical data using Eq.~(\ref{eq:ED-BIC_CDOS}) with fitting parameters $\beta = 44.80\%$, $k_p=0.581\,$rad/$\mu$m for the ED-BIC, and $\beta = 81.79\%$, $k_p=0.562\,$rad/$\mu$m for the MD-BIC. The fits are made over the range $d\geq 5a$ for the ED-BIC and $d\geq 2a$ for the MD-BIC, for which $\Gamma_{12}$ is well-approximated by Eq.~(\ref{eq:ED-BIC_CDOS}) (see SM, Sec.~VI \cite{SM}). The dashed black curves correspond to the free space case, for cross-sections indicated by the white dashed lines in Figs.~\ref{fig:heat-maps}(a,b).} \label{fig:collective_decay}
\end{figure*}

We now consider the coupling between two QEs. Beginning with free space, Figs.~\ref{fig:heat-maps}(a) and (b) show the  spatial dependence of the normalized collective decay rate $\Gamma_{12}/\Gamma_0$ (i.e.~the CDOS), when QE1 is fixed (red dot) and QE2 is scanned across the surface. 
In Fig.~\ref{fig:heat-maps}(a) both dipoles are oriented along $z$, while in Fig.~\ref{fig:heat-maps}(b) they are oriented along $y$.
In order to directly compare with the CDOS in the presence of the metasurface, the transition wavelength is chosen to match the ED-BIC ($\lambda_0=552.0\,$nm) in Fig.~\ref{fig:heat-maps}(a) and the MD-BIC ($\lambda_0=708.9\,$nm) in Fig.~\ref{fig:heat-maps}(b). 
The CDOS in free space is calculated analytically from Eq.~(\ref{eq:gamma_coll}) using the free-space Green’s tensor (see SM, Sec.~IV \cite{SM}).
As expected, in 3D, coupling between two point emitters is limited to close distances ($d<\lambda_0$) because of the isotropic spread of the radiated energy into free space. 


In stark contrast, with the introduction of the metasurface, spatial interactions are completely redefined, as shown in Figs.~\ref{fig:heat-maps}(c) and (d). Here, the position and orientation of QE1 are chosen to maximise coupling to the BIC, as in Figs.~\ref{fig:EDBIC_LDOS_vs_wl}(a) and (b). Remarkably, there is a significant enhancement of the CDOS whenever QE2 is located at nanosphere positions equivalent to that of QE1 within the unit cell. 
Coupling through the BIC modes yields large CDOS values over separations beyond the transition wavelength, $d>\lambda_0$, even exceeding several wavelengths, as we shall see later. This result is rather counter-intuitive, since the BIC system corresponds in principle to an isotropic 2D system, hence the radiation should be spread more or less uniformly, but in a 2D plane instead of 3D space. We will discuss later why the BIC system behaves effectively like a 1D system.

While in these plots QE2 has the same orientation as QE1, interestingly in the MD-BIC case two QEs with orthogonal in-plane orientations (e.g., QE2 aligned along $x$ while QE1 is aligned along $y$) can still couple strongly (see SM, Sec.~V \cite{SM}). 
In fact, for the MD-BIC, the CDOS associated with any in-plane orientation of QE2 can be generated by summing dipole moment magnitude weighted heatmaps of the $x$- and $y$-oriented dipoles. 
This provides greater flexibility than the ED-BIC, where orthogonal orientations do not couple. 
Note that for the MD-BIC, we plot the absolute value of the collective decay rate to improve visual contrast; the full signed $\Gamma_{12}$ distribution is shown in SM, Sec.~V \cite{SM}. For the cross-section considered in later analysis (along the $y$-axis at $x = 0.163a$), $\Gamma_{12}$ is positive, so taking the absolute value does not alter subsequent results.

\begin{table*}[!ht]
            \centering	
            \begin{tabular}{|p{3cm}|p{1.5cm}|p{1.5cm}|p{1.5cm}|p{1.5cm}|p{1.5cm}|}
                \hline
                & $c_0$ & $c_1$ & $c_2$ & $c_3$ & $c_4$\\
                \hline
                ED-BIC  & $0.273$ & $0.516$ & $0.160$ & $0.048$ & $0.001$\\
                \hline
                MD-BIC  & $0.642$ & $0.351$ & $0.005$ & $-$ & $-$\\
                \hline
            \end{tabular}
            \caption{Coefficients $c_n$ of the $(N + 1)$-term Fourier cosine series used in Eq.~(\ref{eq:ED-BIC_CDOS}). For the ED-BIC, $N=4$, and for the MD-BIC, $N=2$.}
            \label{tab0}
        \end{table*}

\subsection{QE-BIC $\beta$-factors}

To further evidence the peculiar nature of the 2D BIC system, we show in Figs.~\ref{fig:collective_decay}(a) and (b) cross-sections of the CDOS for the ED-BIC and MD-BIC, taken along the white dashed lines in Figs.~\ref{fig:heat-maps} (c) and (d), respectively. In the ED-BIC case the separation is  along $x$, while for the MD-BIC it is along $y$. 
As the emitter separation increases, the CDOS exhibits oscillations with gradually decaying envelopes, which are different in the two cases. Interestingly, the decay is not due to a spread of the energy radiated by QE1 in 2D (as it would be in a classical picture), but is primarily a finite-size effect. This is confirmed by comparison with the \emph{infinite} metasurface case (see SM, Sec.~VI \cite{SM}), where the coupling between the two point emitters does not decay with distance, in a similar way as for a (lossless) 1D system such as a waveguide \cite{asenjo2017atom}. Therefore, the BIC system behaves effectively as a 1D system when mediating interactions between distant emitters, but presents the inherent advantages of a 2D system, such as being compatible with 2D arrays of emitters. For reference, the corresponding free space results (in 3D) are also shown (dashed black curves).

To gain further insight, we analyze an infinite metasurface, where closed-form expressions for $\Gamma_{12}$ can be derived by expressing the CDOS 
in terms of an integral over the first Brillouin zone. Using a single-mode approximation for the metasurface, we obtain (see SM, Sec.~VI \cite{SM})
\begin{align}
    &\Gamma_{12}\left(d\right) = \beta\, \Gamma_{11}\, {\rm osc}(d)\,\text{env($d$)},
    \label{eq:ED-BIC_CDOS}
\end{align}
with oscillatory part
\begin{align}
    {\rm osc}(d)=\sum_{n=0}^N c_n \cos\left(2n\pi {d\over a}\right),
    \label{eq:ED-BIC_CDOS_osc}
\end{align}
and envelope defined in terms of a Bessel function of the first kind
\begin{align}
    \text{env($d$)}=J_0\left(k_\parallel^\text{res}|d|\right),
    \label{eq:ED-BIC_CDOS_env}
\end{align}
where $\Gamma_{11}$ is the single-emitter decay rate, $\beta \in [0,1]$ is the $\beta$-factor, defined as the ratio of the emission rate into the BIC mode to $\Gamma_{11}$ \cite{lecamp2007very,manga2007single,arcari2014near}. The oscillatory part 
is an $(N+1)$-term Fourier cosine series, whose expansion coefficients are given in Table~\ref{tab0}. Note that these coefficients are specific to the metasurface parameters 
and QE configuration considered, and would need to be recalculated for a different parameters and configuration. 
Within the envelope,  
the parameter $k_\parallel^\text{res}$ 
quantifies the detuning between the QE transition $\omega_0$ and the BIC resonance, vanishing for perfect frequency matching ($k_\parallel^\text{res}=0$), which then yields ${\rm env}(d)=1$. 
As shown in SM, Sec.~VI \cite{SM}, very good agreement is found between numerical simulations and Eq.~(\ref{eq:ED-BIC_CDOS}) for the infinite metasurface 
\footnote{Numerical results for the infinite metasurface are obtained using a coupled electric and magnetic dipole (CEMD) code developed in Refs.~\cite{Abujetas20,Abujetas21} (see SM, Sec.~I \cite{SM}), because SMUTHI is not applicable.}. 
In this case, the fitted $\beta$-factors 
are $\beta=75.18\%$ for the ED-BIC and $\beta=82.43\%$ for the MD-BIC, which are remarkably high for a 2D platform. 

In a real (finite) metasurface the BIC resonance broadens into a quasi-BIC, effectively introducing detuning from $\lambda_0$. 
This can be accounted for within Eq.~(\ref{eq:ED-BIC_CDOS}) by allowing a non-zero value for $k_\parallel^\text{res}$ to quantify the decay of $\Gamma_{12}$ with distance, which is treated as a fitting parameter. 
In general, the larger (smaller) the metasurface, the smaller (larger) the value of $k_\parallel^\text{res}$. Note also that if weak material absorption were present, it could similarly be accounted for by a non-zero value of $k_\parallel^\text{res}$.
For the fits shown in Figs.~\ref{fig:collective_decay}(a,b) 
($21\times 21$ nanoparticles) 
we obtain $\beta=44.80\%$ with $k_\parallel^\text{res}=0.581\,$rad/$\mu$m for the ED-BIC, and $\beta=81.79\%$ with $k_\parallel^\text{res}=0.562\,$rad/$\mu$m for the MD-BIC. 
Agreement between numerical results and Eq.~(\ref{eq:ED-BIC_CDOS}) is good for both BIC types, especially for larger separations ($d\geq 5a$ for the ED-BIC and $d\geq 2a$ for the MD-BIC), where the BIC mode dominates and the single-mode approximation is accurate. At shorter distances, non-resonant modes contribute more strongly, reducing the approximation’s accuracy. The non-resonant contribution is shown in  SM, Sec.~VI \cite{SM}.

Finite size effects thus reduce $\beta$ for the ED-BIC, but it remains remarkably high for the MD-BIC, comparable to values calculated for photonic crystal waveguides 
\cite{lecamp2007very,manga2007single}. Importantly, however, BIC modes are delocalized in 2D, making metasurfaces much more readily integrable with large ensembles of QEs in 2D arrays. Moreover, while for comparison with existing literature we have chosen metasurface parameters identical to those in Ref.~\cite{Abujetas21}, optimisation of the metasurface geometry and meta-atom type (e.g., nano-disks, rings) could significantly increase $\beta$ values.

Although finite size also introduces decay of the long-range interaction ($k_\parallel^\text{res}\neq 0$), coupling persists across several micrometers for both BIC types, representing a substantial enhancement compared to free space. While the ED-BIC yields higher LDOS and CDOS, its $\beta$-factor is smaller. As we show in the next section, the $\beta$-factor is critical in determining entanglement generation over remote distances, which is made possible by BIC modes. 

\subsection{Concurrence} 

\begin{figure*}
       \centering
        \includegraphics[width=0.8\linewidth]{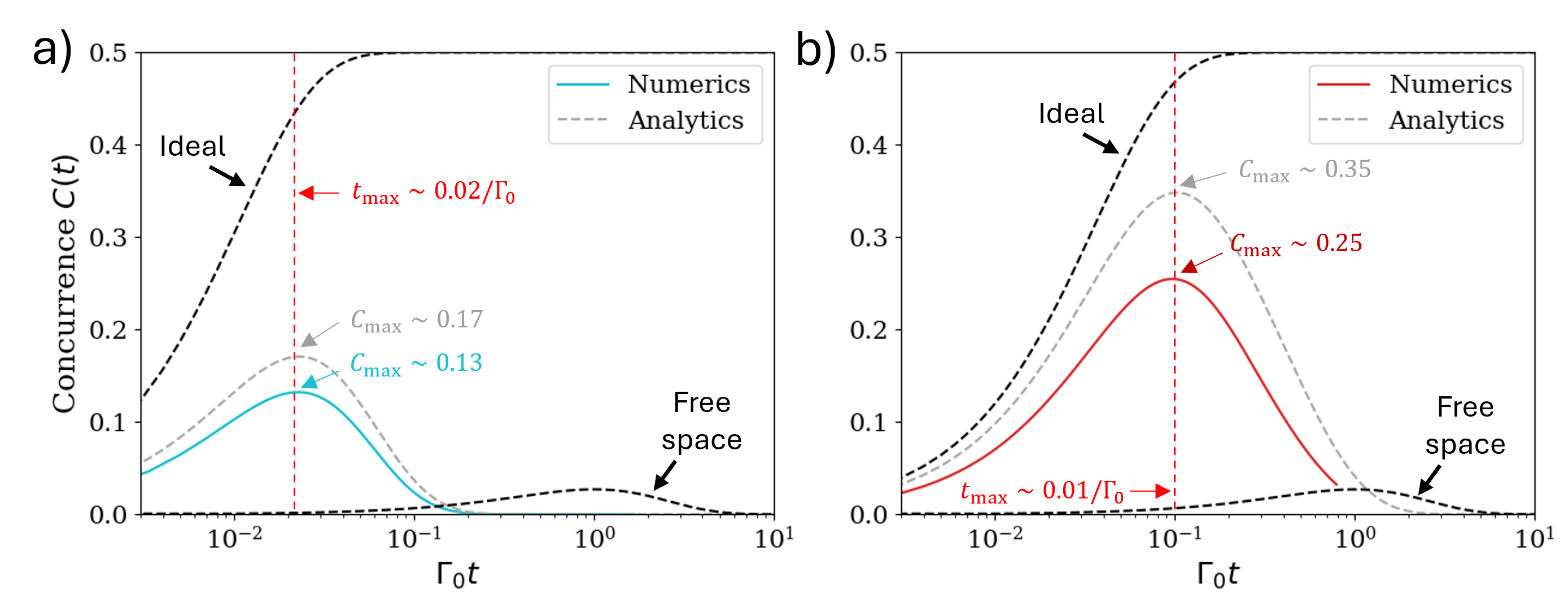} 
        \caption{
        Concurrence $C(t)$ as a function of dimensionless time $\Gamma_0t$ for a QE separation of $d=2\,\mu$m in the case of the ED-BIC (a) and the MD-BIC (b). Solid curves are found numerically, while the dashed grey curves are obtained from Eq.~(\ref{eq:conc_approx}) with $\beta=44.80\%$, $F_p=46.9$ for the ED-BIC and $\beta=81.79\%$, $F_p=13.7$ for the MD-BIC. Ideal metasurface coupling cases ($\beta=1$) and the concurrence in free space for similar separations 
        [$d=1.79\,\mu$m in (a) and $d=2.30\,\mu$m in (b), chosen to match CDOS free space maxima] are also shown.} 
        \label{fig:conc_vs_t}
\end{figure*}

\begin{table*}[!ht]
            \centering	
            \begin{tabular}{ |p{3cm}|p{1.5cm}|p{1.5cm}|p{1.5cm}|p{1.5cm}|}
                \hline
                & $\Gamma_{11}/\Gamma_0$ & $\Gamma_{22}/\Gamma_0$ & $\Gamma_{12}/\Gamma_0$ & $\Omega_{12}/\Gamma_0$\\
                \hline
                ED-BIC  & $46.9$ & $42.4$ & $15.3$ & $-2.6$\\
                \hline
                MD-BIC  & $13.7$ & $8.8$ & $7.9$ & $-0.2$\\
                \hline
            \end{tabular}
            \caption{Master equation coefficients used in the numerical computation of the concurrence in Figs.~\ref{fig:conc_vs_t} (a) and (b). The rates $\Gamma_{11}$ are extracted from Figs.~\ref{fig:EDBIC_LDOS_vs_wl} (a) and (b) at $\lambda_0=552.0\,$nm and $\lambda_0=708.9\,$nm, respectively, and $\Gamma_{12}$ from Figs.~\ref{fig:collective_decay} (a) and (b) at $d=5a$ ($2\,\mu$m). $\Gamma_{22}$ and $\Omega_{12}$ are extracted from \textcolor{blue}{Fig.~S4} at $d=5a$ ($2\,\mu$m).}
            \label{tab1}
        \end{table*}

As an important application of the above results, we now demonstrate that metasurfaces can generate long-range QE–QE entanglement. To quantify entanglement, we use the concurrence \cite{Wootters1998,Wootters2001}
\begin{equation}
    C(t) = \max(0, \sqrt{\lambda_1} -  \sqrt{\lambda_2}  - \sqrt{\lambda_3}  - \sqrt{\lambda_4}),
    \label{eq:conc0}
\end{equation}
where $\lambda_i$ are the eigenvalues (ordered by decreasing magnitude) of the non-Hermitian matrix $R = \rho \Tilde{\rho}$, with  $\Tilde{\rho} = (\sigma_y \otimes \sigma_y) \rho^* (\sigma_y \otimes \sigma_y)$, in which conjugation is taken in the Pauli-$Z$ basis. The concurrence satisfies $0\leq C(t) \leq 1$, with $C = 0$ for separable states and $C = 1$ for maximally entangled states.

We consider an initially uncorrelated QE state $\ket{\mathrm{e}_1\mathrm{g}_2}$ in the electromagnetic vacuum. In this case, a closed-form expression for $C(t)$ in terms of the $\beta$-factor can be derived under the following assumptions:
(i) the QEs interact solely via the BIC mode, such that $\Gamma_{12}(d)$ is given by Eq.~(\ref{eq:ED-BIC_CDOS}) (single-mode approximation); (ii) the two QEs have the same LDOS at positions separated by integer multiples of the lattice constant, $d=na$ with $n$ an integer, so that $\Gamma_{22}= \Gamma_{11}=F_p\Gamma_0$ (which is exact for QEs interacting with an infinite metasurface);  
(iii) the interaction is purely dissipative, that is, $\Omega_{12}= 0$. 

Upon imposing approximations (i)-(iii), and at separations $d=na$, the concurrence becomes: 
(see SM, Sec.~II \cite{SM})
\begin{equation}
C(t)\simeq \mathrm{sinh}\left(\beta F_p\Gamma_0t\right)\mathrm{e}^{-F_p\Gamma_0t}
\label{eq:conc_approx}
\end{equation}
An important consequence is that the concurrence becomes independent of the QE separation: as long as the emitters occupy equivalent positions with respect to the unit cell, the degree of entanglement is the same, regardless of $d$.


Figures~\ref{fig:conc_vs_t}(a) and (b) compare the analytical predictions of Eq.~(\ref{eq:conc_approx}) (dashed grey curves) with numerical results (solid curves) for the concurrence evolution at $d=5a=2\,\mu$m. The numerical results are obtained by solving the two-QE master equation for a $21\times 21$ metasurface (see SM, Sec.~I \cite{SM}). We see that entanglement develops for both BIC
types and that the agreement between the numerical and analytical curves is qualitatively good, though finite-size effects mean approximation (ii) is not exactly satisfied (see Table~\ref{tab1}). The discrepancy is larger for the MD-BIC, where $\Gamma_{11}\neq\Gamma_{22}$ is more pronounced. Note that approximation (iii) is relatively well satisfied (see Table~\ref{tab1}), and this is generally the case for for $d>\lambda_0$  (see SM, Sec.~VII \cite{SM}). 
For comparison, the free-space case is also shown, where $\Gamma_{11}=\Gamma_{22}=\Gamma_0$ and $\Omega_{12}\rightarrow0$ for the separations chosen, such that Eq.~(\ref{eq:conc_approx}) becomes exact \footnote{In free space $\Gamma_{12}$ and $\Omega_{12}$ are in quadrature, i.e. $\pi/2$ out-of-phase, see SM, Sec.~IV \cite{SM}.}. Entanglement in free space develops at least an order of magnitude later and with much smaller magnitude than in the metasurface platform.

        
\begin{figure*}
       \centering
        \includegraphics[width=0.8\linewidth]{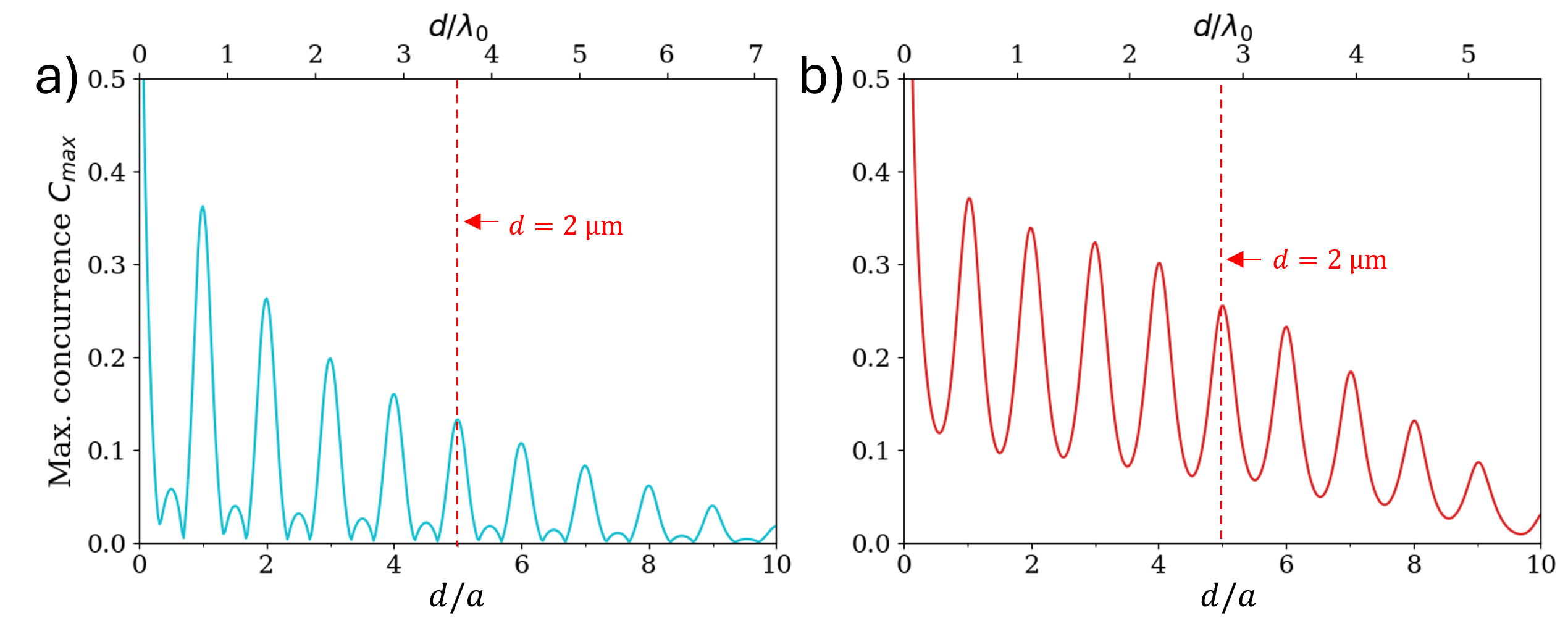} 
        \caption{
        Maximum concurrence $C_\text{max}$ plotted with emitter separation $d$ for the ED-BIC (a) and MD-BIC (b). Other parameters are as in Fig.~\ref{fig:conc_vs_t}.} \label{fig:conc_vs_d}
\end{figure*}

At long times, Eq.~(\ref{eq:conc_approx}), converges to the population of the asymmetric Dicke state $\rho_{aa}(t)$ (see SM, Sec.~II \cite{SM}), and the entanglement eventually decays exponentially as
\begin{equation}
C(t)\simeq \frac{1}{2}\mathrm{e}^{-(1-\beta) F_p\Gamma_0 t}.
\label{eq:conc_long_time_limit}
\end{equation}
Thus, smaller Purcell factors $F_p$ correspond to longer-lived entanglement, which explains why the 
MD-BIC ($F_p\sim 14$) sustains entanglement longer than the ED-BIC ($F_p\sim 47$). 
Note that the ED-BIC and MD-BIC also possess different $\beta$. 
As $\beta \to 1$ the exponential decay is increasingly suppressed and in the ideal case $C(t)\to 0.5$ for large $t$. 
We plot this for both BICs (corresponding to a metasurface with $\beta=1$, keeping $F_p$ fixed) in Fig.~\ref{fig:conc_vs_t}.  

We see that for the ED-BIC, the maximum concurrence is smaller, occurs sooner, and decays faster than for the MD-BIC.
This can be understood by noting that the maximum occurs at
\begin{equation}
t_\mathrm{max} = \frac{1}{2F_p\Gamma_0\beta}\mathrm{ln}\left(\frac{1+\beta}{1-\beta}\right),
\label{eq:t_max}
\end{equation}
with value
\begin{equation}
C_\mathrm{max}=\frac{\beta}{\sqrt{1-\beta^2}}\left(\frac{1+\beta}{1-\beta}\right)^{-\frac{1}{2\beta}}.
\label{eq:C_max}
\end{equation}
This shows importantly that $C_\mathrm{max}$ is a monotonically increasing function of $\beta$ alone, which is why it is larger in the case of the MD-BIC. Conversely, the time to reach maximum entanglement is set by both $\beta$ and $F_P$, so the ED-BIC develops entanglement sooner, while the MD-BIC sustains it longer.
 

Finally, Fig.~\ref{fig:conc_vs_d} shows the numerically computed $C_\text{max}$ as a function of emitter separation $d$. The envelope follows the decay of the collective rates (Fig.~\ref{fig:collective_decay}), with oscillations that would thus persist indefinitely in the infinite-metasurface limit (see SM, Sec.~VI \cite{SM}). At short separations, additional non-resonant modes enhance entanglement in the ED-BIC, while in the MD-BIC case significant enhancement appears for subwavelength separations $d < \lambda_0$, where dipole-dipole interactions become significant.



\section{Conclusions}

We have shown that quasi-BICs in non-local (finite) metasurfaces efficiently mediate long-range interactions between QEs placed in their near-field (here, $4\,$nm above the metasurface). The ED-BIC couples to out-of-plane dipoles, while the MD-BIC couples to in-plane dipoles. In the weak-coupling and dissipative regimes, their interactions are governed by the Purcell factor $F_p$ and the coupling efficiency $\beta$. 

While $F_p$ can be obtained numerically, we extracted $\beta$-factors from fits of CDOS simulations to an analytical expression derived for infinite metasurfaces [Eq.~(\ref{eq:ED-BIC_CDOS})].
For the ED-BIC we found $F_p \sim 47$ and $\beta \sim 0.45$, while for the MD-BIC we obtained $F_p \sim 14$ and $\beta \sim 0.82$. In the latter case, the $\beta$-factor is remarkably high and comparable to those reported for one-dimensional waveguides \cite{manga2007single,lecamp2007very}.

As a key application, we studied entanglement generation between two QEs initially prepared in the state $\ket{\mathrm{e}_1, \mathrm{g}_2}$ (in the electromagnetic vacuum). We showed that the maximum concurrence $C_\text{max}$ depends primarily on $\beta$, increasing monotonically with it. Since the MD-BIC possesses a higher $\beta$-factor, it yields stronger entanglement ($C_\text{max} \sim 0.25$) than the ED-BIC ($C_\text{max} \sim 0.13$) at a QE separation of $2\,\mu$m ($\simeq 3.6\lambda_0$ for the ED-BIC and $\simeq 2.8\lambda_0$ for the MD-BIC). Larger metasurfaces would further enhance performance, approaching the infinite limit predicted by Eqs.~(\ref{eq:conc_approx}) and (\ref{eq:C_max}). This demonstrates that non-local metasurfaces can generate entanglement through purely dissipative interactions, even across distances of several wavelengths.

While these values of concurrence are not necessarily larger than those computed in other systems \cite{gonzalez2011entanglement,Jha2017}, they can be improved by optimizing the metasurface design to engineer higher $\beta$-factors; for example, by tailoring nanoparticle shapes to better isolate BIC resonances over a broader spectral range. 
Crucially, while the entanglement reported here spans separations of several wavelengths, it could in principle persist over arbitrarilyy larger distances in larger metasurfaces, since according to Eqs.~(\ref{eq:conc_approx}) and (\ref{eq:C_max}) it is independent of the separation $d$, provided $d \ll c/\Gamma_{11}$ (beyond which retardation effects become significant \cite{kastel2005suppression}. 

We also found that entanglement develops faster for the ED-BIC but decays more rapidly, since the time of maximum entanglement scales inversely with $F_p$, while its long-time decay rate scales with $F_p(1-\beta)$ [Eqs.~(\ref{eq:conc_long_time_limit}) and (\ref{eq:t_max}), respectively]. In the ideal limit $\beta = 1$, the concurrence plateaus at $C_\text{max} =0.5$.

Although we focused here on two emitters, the same principles extend to larger ensembles. The framework naturally generalizes to many-body systems ($N > 2$ emitters) with long-range, all-to-all interactions mediated by metasurfaces \cite{Alvarez2020, Gardiner2022}. Similarly, while we modeled QEs as ideal two-level systems, the formalism can be readily adapted to multi-level emitters and to include dephasing.

Looking forward, non-local metasurfaces offer a practical route to coupling distant qubits, as they naturally accommodate large 2D arrays of QEs due to their delocalized BIC modes. Their planar geometry and resulting integrability with QE arrays make them especially well-suited for realising scalable quantum technologies. For example, the architecture explored here could enable quantum simulation of spin models with nearest-neighbor or long-range interactions, offering a versatile platform for future quantum nanophotonics.

\section*{Acknowledgements}

E.L. and R.P.-D. acknowledge initial discussions with Dmitry A. Kalashnikov that led to the formulation of this project, and thank Francisco J. Garcia-Vidal for helpful discussions about the $\beta$-factor.
The authors thank Daniel E. Eyles for his help in performing the simulations of the insets in Figures 2 (c)-(d) using SMUTHI.
H.R. acknowledges the Turing scheme for supporting her stay in Singapore.
E.L. and R.P.-D. acknowledge funding support from Singapore MTC-Programmatic Grant No. M21J9b0085.


\bibliography{biblio_file}

\newpage 

\include{supplementary_information.tex}

\end{document}

%% file: supplementary_information.tex





\clearpage
\widetext
\begin{center}
\textbf{\large Supplemental Material: \\Long-range quantum emitter interactions mediated by a non-local metasurface: Application to qubit-qubit entanglement}
\end{center}






\setcounter{section}{0}

\section{\label{app:methods} Methods}

\subsection{SMUTHI} 

We use the Python package SMUTHI~\cite{egel2021smuthi} for the electromagnetic simulations of the coefficients $\Gamma_{11}$, $\Gamma_{22}$, $\Gamma_{12}$, and $\Omega_{12}$. In all simulations, we make the electric and magnetic dipole approximations to model the metasurface, which involves setting the multipole order $l_\text{max}=1$ in SMUTHI. This helps to speed-up simulations considerably, and is sufficiently accurate to capture the phenomena associated to the optical resonances of interest.


The LDOS $\Gamma_{\mu\mu}/\Gamma_0$ is obtained by computing the left-hand-side of the relation $P/P_0 = \Gamma_{\mu\mu}/\Gamma_0$. The power dissipated by the dipole $\mu$ in the presence of the metasurface, $P$, and the corresponding free-space power $P_0$, are calculated using SMUTHI.

The CDOS $\Gamma_{12}/\Gamma_0$ is obtained by expressing the dyadic Green's tensor in terms of the electric field scattered by the metasurface, which is found using SMUTHI. 
To construct the heatmaps (Fig.~3), we sweep our observation point (the target dipole) over the entire metasurface, in the $xy$-plane located at $z=R+h$ above the metasurface, to obtain the scattered electric field at each point, and from this, the associated collective decay rate.\\

\subsection{CEMD} 

The coupled electric and magnetic dipole (CEMD) model developed in Refs.~\cite{Abujetas20, Abujetas21} was used to calculate the collective decay rate for the infinite metasurface cases. For this purpose the self-consistent problem of the scattering of 
the electric source field of a dipole by an 
infinite metasurface is solved, where in the numerical implementation, the dipole source field is expanded in plane waves. 
The polarizabilities of the spherical particles are calculated using Mie theory.\\


\subsection{QuTIP} 

We use the Quantum Toolbox in Python (QuTIP)~\cite{JOHANSSON20121760,lambert2024qutip5quantumtoolbox} to solve the master equation given in Section~\ref{app:ME}. The input master equation coefficients $\Gamma_{\mu\nu}$ and $\Omega_{\mu\nu}$ are the outputs of the SMUTHI computations. We assume a real transition dipole moment for both QEs with magnitude $p=1\times10^{-29}\,$Cm, which is $\sim 3\,$Debye.

\section{\label{app:ME} Quantum master equation formalism}

\subsection{System Hamiltonian}

We begin with the Hamiltonian describing two identical two-level dipoles ($\mu=1,2$) comprising an atomic (quantum emitter) system which is embedded within a dielectric medium \cite{dung2002resonant}
\begin{align}
    H &= H_A + H_M + V\nonumber\\
    &= \sum_{\mu=1}^2\frac{1}{2}\hbar \omega_0 \sigma^z_{\mu} + \int d^3\bm{r} \int_0^{\infty}d \omega \hbar \omega \bm{f}^{\dagger}(\bm{r},\omega) \bm{f}(\bm{r},\omega) + \sum_{\mu=1}^2(\sigma^+_{\mu} + \sigma^-_{\mu}) \bm{d} \cdot \bm{E}_M(\bm{R}_{\mu}).
\end{align}
This Hamiltonian is given in the dipole gauge \cite{stokes2018master,Stokes2022} 
with the atomic part expressed in terms of the Pauli spin operators $\sigma_\mu^z$ and the transition frequency $\omega_0$, which corresponds to the transition from the ground to the excited state, $\ket{\mathrm{g}} \rightarrow \ket{\mathrm{e}}$. The medium Hamiltonian encompasses the total environment which is made up of the dielectric medium and free space. The operator-valued bosonic field variables $f_i(\bm{r},\omega)$ and $f_i^{\dagger}(\bm{r},\omega)$ respresent the elementary excitations of the total environment. The interaction Hamiltonian $V$ is decomposed in terms of Hermitian system and environment operators. 
The electric field operators are defined in terms of the continuum of bosonic field operators and the respective dipole position $\bm{R}_{\mu}$ as \cite{dung2002resonant}
\begin{align}\label{pi_components}
    E_{M_j}^+(\bm{R}_{\mu}) &=  i \sqrt{\frac{\hbar }{\pi \epsilon_0}}\frac{1}{c^2} \int d^3\bm{r'}\int_0^{\infty} d \omega \omega^2 \sqrt{\epsilon_I(\bm{r'},\omega)} G_{ji}(\bm{R}_{\mu}, \bm{r'}, \omega) f_i(\bm{r'},\omega),\\
    E_{M_j}^-(\bm{R}_{\mu}) &=  - i \sqrt{\frac{\hbar}{\pi \epsilon_0}}\frac{1}{c^2} \int d^3\bm{r'}\int_0^{\infty} d \omega \omega^2  \sqrt{\epsilon_I(\bm{r'},\omega)} G_{ji}(\bm{R}_{\mu}, \bm{r'}, \omega) f_i^{\dagger}(\bm{r'},\omega).
\end{align}
where $\epsilon_I(\bm{r},\omega) = 1 + \int_0^{\infty}d\tau \chi(\bm{r},t)e^{i \omega \tau}$ is the relative permittivity with dielectric susceptibility $\chi(\bm{r},t)$, and $G_{ij}(\bm{r},\bm{r'},\omega)$ is the dyadic Green's tensor. The dyadic Green's tensor governs the properties of the dielectric medium \cite{Knoll2003} and is key to bridging the gap between our quantum and classical nanophotonic formalisms. Note that there is an implicit summation over the cartesian coordinates $i$ above and in the following. 

\subsection{Master equation}

We adopt a standard open systems approach to derive an equation of motion for the reduced density matrix of the two-emitter system \cite{Breuer2007}. After a lengthy but straightforward calculation, we arrive at the secular Born-Markov master equation, 
\begin{align}\label{eqn:secular_born_markov_master_equation}
    \dot{\rho}_S(t) =& -i \sum_{\mu=\nu=1}^2\Tilde{\omega}_0[\sigma_{\mu}^+\sigma_{\nu}^-, \rho_S(t)] - i\sum_{\mu \neq \nu}^2\Omega_{\mu\nu}[\sigma_{\mu}^+\sigma_{\nu}^-, \rho_S(t)] \nonumber\\ &+ \sum_{\mu,\nu = 1}^2 \Gamma_{\mu\nu}(\omega_0)\left(\sigma_{\mu}^-\rho_S(t)\sigma_{\nu}^+ - \frac{1}{2}\{\sigma_{\nu}^+\sigma_{\mu}^-, \rho_S(t)\}\right),
\end{align}
This equation is analogous to the master equation for two dipoles in free space \cite{Agarwal2012,stokes2018master}, with differences due to the presence of the dielectric medium contained within the coefficients \cite{dung2002resonant}.

Specifically, we have
\begin{align}\label{eqn:total_coupling}
    \Omega_{\mu \nu}(\bm{R}_{\mu},\bm{R}_{\nu},\omega_0)    = \frac{ 2\omega_0}{ \hbar \pi \epsilon_0 c^2}\mathcal{P}\int_0^{\infty}\text{d}\omega d_i d_j   \Im(G_{ij}(\bm{R}_{\mu},\bm{R}_{\nu},\omega_0)) \frac{\omega^2}{\omega_0^2 - \omega^2}.
\end{align}
where the symbol $\mathcal{P}$ denotes the principal value of the integral and the diagonal elements correspond to the (Lamb shift) self-energy interactions that modify the atomic transition frequencies, 
\begin{equation}\label{Tildeomega0}
\Tilde{\omega}_0 = \omega_0 + \Omega_{\mu = \nu},
\end{equation}
and the off-diagonal elements correspond to the collective coupling rate 
\begin{equation}\label{eqn:collective_coupling_rate}
\Omega_{\mu\neq\nu} = - \frac{\omega_0^2}{\hbar \varepsilon_0 c^2} d_i d_j \Re(G_{ij}(\bm{R}_{\mu},\bm{R}_{\nu},\omega_0)).
\end{equation}
The total decay rate is defined as
\begin{equation}\label{eqn:total_decay_rate}
\Gamma_{\mu \nu} (\bm{R}_{\mu},\bm{R}_{\nu},\omega_0) = \frac{2  d_i d_j  \omega_0^2}{ \hbar \epsilon_0 c^2}  \Im(G_{ij}(\bm{R}_{\mu},\bm{R}_{\nu},\omega_0)).
\end{equation}
We have taken the zero temperature limit such that the bosonic occupation number $N=0$. 

The first term in equation Eq.~\eqref{eqn:secular_born_markov_master_equation} describes the coherent, unitary evolution of the system and  
is defined in the same way as 
in the single dipole case. The second term pertains to the off-diagonal dipole-dipole shifts with coherent collective coupling rate $\Omega_{\mu \neq \nu}$. These shifts contain spatial dependence arising from the Green's function and have no analogue in the single dipole case. In order to calculate $\Omega_{\mu \neq \nu}$, an integral over the entire spectrum must be performed. 
The integral can be evaluated 
through the introduction of an upper-frequency cut-off. Alternatively, using properties of the Green's function it is possible to simplify the calculation significantly. The full dyadic Green's function $G_{ij}(\bm{R}_{\mu},\bm{R}_{\nu},\omega)$ has no poles in the upper complex half-plane due to causality. This means that a Kramers-Kronig relation holds. When one takes the zero temperature limit ($N=0$) of the dipole-dipole shift integral in Eq.~\eqref{eqn:total_coupling}, this relation may be used to yield the result in Eq.~\eqref{eqn:collective_coupling_rate}, which is given in terms of the real part of the full Green's function \cite{Dzsotjan2011}. 

The final term in Eq.~\eqref{eqn:secular_born_markov_master_equation} describes dissipative emission processes with the associated rates $\Gamma_{\mu\nu}(\omega_0)$. Note that as we have taken the zero temperature limit, there are no absorption processes. The diagonal matrix elements yield the spontaneous (single-dipole) decay rates. 
The off-diagonal matrix elements correspond to the collective (inter-dipole) decay rates, which contain spatial dependence in the same way as the collective coupling rate. The decay rates are given in terms of the imaginary part of the full Green's function as seen in Eq.~\eqref{eqn:total_decay_rate}.  We remark that in deriving Eq.~\eqref{eqn:secular_born_markov_master_equation} we do not apply a rotating-wave approximation to the Hamiltonian, but instead perform a secular approximation to remove fast oscillating 
terms at the master equation level.

\subsection{Solutions of the master equation}

To solve the master equation 
we move from the computational basis to the collective Dicke basis $\{\ket{\mathrm{e}} = \ket{\mathrm{ee}}, \ket{\mathrm{s}}=\frac{1}{\sqrt{2}}(\ket{\mathrm{eg}} + \ket{\mathrm{ge}}), \ket{\mathrm{a}}=\frac{1}{\sqrt{2}}(\ket{\mathrm{eg}} - \ket{\mathrm{ge}}), \ket{\mathrm{g}} = \ket{\mathrm{gg}}\}$ as this best elicits the quantum nature of the interactions. 
The resulting differential equations for the Dicke basis atomic populations are given by 
\begin{align}
    \dot{\rho}_{ee}(t) &= -2\Gamma \rho_{ee}(t),\\
    \dot{\rho}_{ss}(t) &= -(\Gamma + \Gamma_{12})\rho_{ss}(t) - \frac{(\Gamma_{11} - \Gamma_{22})}{4} \biggl(\rho_{as}(t) + \rho_{sa}(t)\biggr) + (\Gamma + \Gamma_{12}) \rho_{ee}(t),\\
    \dot{\rho}_{aa}(t) &= - (\Gamma - \Gamma_{12}) \rho_{aa}(t) - \frac{(\Gamma_{11} - \Gamma_{22})}{4} \biggl(\rho_{as}(t) + \rho_{sa}(t) \biggr) + (\Gamma - \Gamma_{12}) \rho_{ee}(t),
\end{align}
with the condition $\rho_{gg}(t) = 1 - \rho_{ee}(t) - \rho_{ss}(t) - \rho_{aa}(t)$, and for the relevant coherences we have
\begin{align}
    \dot{\rho}_{as}(t) &= - (\Gamma - 2i\Omega_{12}) \rho_{as}(t) - \frac{(\Gamma_{11} - \Gamma_{22})}{4}(\rho_{ss}(t) + \rho_{aa}(t) + 2\rho_{ee}(t)),\\
    \dot{\rho}_{sa}(t) &= - (\Gamma + 2i\Omega_{12})\rho_{sa}(t) - \frac{(\Gamma_{11} - \Gamma_{22})}{4}(\rho_{ss}(t) + \rho_{aa}(t) + 2\rho_{ee}(t)),
\end{align}
where $\Gamma = {(\Gamma_{11} + \Gamma_{22})}/{2}$ is the average of the two individual dipole decay rates 
and we have used $\Gamma_{12} = \Gamma_{21}$. 


In general, the solutions to the above differential equations do not given much analytical insight, and so we solve them numerically for the corresponding plots given in the main text and elsewhere in this supporting information. However, in cases where $\Gamma_{11} = \Gamma_{22}$, as assumed for the approximate analytic concurrence expression given in the main text, 
the populations and coherences completely decouple. This leads to simplified differential equations 
\begin{align}
    \dot{\rho}_{ee}(t) &= -2\Gamma \rho_{ee}(t),\\
    \dot{\rho}_{aa}(t) &= - (\Gamma - \Gamma_{12}) \rho_{aa}(t) + (\Gamma - \Gamma_{12}) \rho_{ee}(t),\\
    \dot{\rho}_{ss}(t) &= -(\Gamma + \Gamma_{12})\rho_{ss}(t) + (\Gamma + \Gamma_{12}) \rho_{ee}(t),\\
     \dot{\rho}_{as}(t) &= - (\Gamma - 2i\Omega_{12}) \rho_{as}(t),\\
    \dot{\rho}_{sa}(t) &= - (\Gamma + 2i\Omega_{12})\rho_{sa}(t),
\end{align}
with the closed form analytical solutions 
\begin{align}
    \rho_{ee}(t) &= \rho_{ee}(0) e^{- 2\Gamma t},\\
    \rho_{ss}(t) &= \rho_{ss}(0) e^{-\Gamma_+ t} + \frac{\Gamma_+}{\Gamma_-} \rho_{ee}(0)\left(  e^{-\Gamma_+ t} - e^{- 2\Gamma t} \right),\\
    \rho_{aa}(t) &= \rho_{aa}(0) e^{-\Gamma_- t} + \frac{\Gamma_-}{\Gamma_+} \rho_{ee}(0)\left(  e^{-\Gamma_- t} - e^{- 2\Gamma t} \right),\\
    \rho_{as}(t) &= \rho_{as}(0) e^{-(\Gamma - 2i\Omega_{12})t},\\
    \rho_{sa}(t) &= \rho_{sa}(0) e^{-(\Gamma + 2i\Omega_{12})t}.
\end{align}
Here we denote $\Gamma_+ = (\Gamma + \Gamma_{12})$ and $\Gamma_- = (\Gamma - \Gamma_{12})$, corresponding to decay rates enhanced (superradiant) and suppressed (subradiant) by collective effects, respectively. 
Such effects are characteristically quantum in nature. These modifications are present in free space for very small dipole separations, however they are significantly amplified by the presence of 
the dielectric medium. It should be noted that our solutions apply to the general case of two emitters in the presence of a medium, one need only to appropriately adjust the the Green's function ansatz and propagate it through. 

\subsection{Concurrence}

In the case considered in the main text where initially only one quantum emitter is excited, 
the concurrence takes the analytical form \cite{tanas2004entangling} 
\begin{equation}
C(t) = \sqrt{[\rho_{ss}(t)-\rho_{aa}(t)]^2-[\rho_{sa}(t)-\rho_{as}(t)]^2}.
\label{eq:conc1}
\end{equation}
We see from Eq.~(\ref{eq:conc1}) 
two limiting cases: (i) $t=0$, for which the initial conditions correspond to $\rho_{ss}(0)=\rho_{aa}(0)=\rho_{as}(0)=1/2$ and $\rho_{ee}(0)=0$, and the concurrence is $C(0)=0$,
since the initial state is unentangled; 
and (ii) $t\rightarrow\infty$, where the concurrence is $C(\infty)\rightarrow 0$ since all excited populations and coherences decay exponentially in time (unless 
$\Gamma_{12}=\Gamma_{11}$, which is satisfied only for unseparated emitters or for a $\beta$-factor equal to unity, where $C(\infty)\rightarrow 0.5$). Therefore, 
entanglement 
is dynamically generated in the \emph{transient} regime. By using the fact that $\rho_{sa}(t) = \rho_{as}^*(t)$,  Eq.~(\ref{eq:conc1}) can be rewritten as 
\begin{equation}
C(t)= \sqrt{[\rho_{ss}(t)-\rho_{aa}(t)]^2-4[\mathrm{Im}(\rho_{as}(t))]^2}.
\end{equation}
Using the analytical expressions for the density matrix elements given above, we then find
\begin{equation}
C(t)=\frac{1}{2}\sqrt{[\mathrm{e}^{-(\Gamma'+\Gamma_{12})t}-\mathrm{e}^{-(\Gamma'-\Gamma_{12})t}]^2+4\mathrm{e}^{-2\Gamma't}\mathrm{sin}^2(2\Omega_{12}t)}.
\label{eq:conc3}
\end{equation}
By making the approximations mentioned in the main text, $\Gamma_{22}\approx\Gamma_{11}$ and $\Omega_{12}\approx 0$, Eq.~(\ref{eq:conc3}) simplifies to:
\begin{equation}
C(t)\simeq \text{sinh}(\Gamma_{12}t)\mathrm{e}^{-\Gamma_{11}t}.
\label{eq:conc_supp_info}
\end{equation}
This takes the form given in Eq.~(6) of the main text once we approximate $\Gamma_{12}\approx \beta\,\Gamma_{11}$ and use the fact that $\Gamma_{11}=F_p\Gamma_0$.

\section{Validity of the weak-coupling approximation}

When an emitter is coupled to a nanophotonic environment, the condition for strong-coupling is usually given by \cite{torma2014strong}:
\begin{equation}
4g^2 > \frac{\Gamma_0^2+\Gamma_\text{BIC}^2}{2}
\end{equation}
where $g$ is the coupling constant, $\Gamma_0$ is the emitter lifetime in free space and $\Gamma_\text{BIC}$ is the lifetime of the mode to which the emitter couples to, which is a BIC mode in our case.
The coupling constant $g$ is related to the Purcell factor $F_p$ as: $F_p=4g^2/(\Gamma_0\Gamma_{BIC})$.
The BIC lifetime is related to the linewidth or full width at half maximum (FWHM) as: $\Gamma_\text{BIC}\simeq 2\pi c\,\text{FWHM}/\lambda_\text{BIC}^2=2\pi c /(\lambda_\text{BIC}Q)$, where we defined the $Q$-factor of the BIC as: $Q=\lambda_\text{BIC}/\text{FWHM}$.
Hence, the strong-coupling condition can be rewritten as:
\begin{equation}
F_p>\frac{\Gamma_0\lambda_\text{BIC}Q}{4\pi c} + \frac{\pi c}{\Gamma_0\lambda_\text{BIC}Q}.
\end{equation}
The first term on the right hand side of the inequality is negligible compared to the second term, so the condition becomes:
\begin{equation}
F_p>\frac{\pi c}{\Gamma_0\lambda_\text{BIC}Q}.
\end{equation}
In our case, for the ED-BIC, one has $F_p\sim 47$, $\lambda_\text{BIC}=552\,$nm, $\text{FWHM}\sim 2\,$nm and $Q\sim 276$ (see Fig.~2 in the main text). By using the expression of the emitter decay rate in free space: $\Gamma_0=\omega_0^3p^2/(3\pi\varepsilon_0\hbar c^3)$, the strong-coupling condition is met for dipole moments $p>p_\text{max}=90\cdot 10^{-29}\,$Cm, which corresponds to approximately $p_\text{max}\sim 270\,$Debye.
Similarly, for the MD-BIC, one has $F_p\sim 14$, $\lambda_\text{BIC}\sim 709\,$nm, $\text{FWHM}\sim 0.05\,$nm and $Q\sim 14\,178$ (see Fig.~2 in the main text). The strong-coupling condition is thus met for dipole moments $p>p_\text{max}=30\cdot 10^{-29}\,$Cm, which corresponds to approximately $p_\text{max}\sim 90\,$Debye. Therefore, the weak-coupling approximation made in this study will generally be valid for emitter dipole moment $p\ll 270\,$Debye in the ED-BIC case, and $p\ll 90\,$Debye in the MD-BIC case.

\section{Green's function in free space}

In free space, the Green's function $\mathbf{G}$ is given by \cite{Agarwal2012}:
\begin{equation}\label{realimagGreens}
    G_{ij}(\bm{R}_{\mu},\bm{R}_{\nu},\omega) = \frac{\omega}{4 \pi c}(- \kappa_{ij}(\bm{R}_{\mu},\bm{R}_{\nu},\omega) 
        + i\tau_{ij}(\bm{R}_{\mu},\bm{R}_{\nu},\omega))\, ,
\end{equation}
where
\begin{equation}
    \kappa_{ij}(\bm{R}_{\mu},\bm{R}_{\nu},\omega) = - (\delta_{ij} - \hat{R}_i\hat{R}_j)\frac{\cos(\theta)}{\theta} + (\delta_{ij} - 3\hat{R}_i\hat{R}_j)\biggl[\frac{\sin(\theta)}{\theta^2} + \frac{\cos(\theta)}{\theta^3}\biggr] ,\\
    \label{eq:re_green}
\end{equation}
and 
\begin{equation}
    \tau_{ij}(\bm{R}_{\mu},\bm{R}_{\nu},\omega) = (\delta_{ij} - \hat{R}_i\hat{R}_j)\frac{\sin(\theta)}{\theta} + (\delta_{ij} - 3\hat{R}_i\hat{R}_j)\biggl[\frac{\cos(\theta)}{\theta^2} - \frac{\sin(\theta)}{\theta^3}\biggr] ,
    \label{eq:im_green}
\end{equation}
with $\theta = \frac{2\pi}{\lambda_0}\frac{R}{c}$, $R$ is the magnitude of $\bm{R} = \bm{R}_{\mu} - \bm{R}_{\nu}$, and $\hat{R_i}$ is the projection of the unit vector 
$\hat{\bm{R}} = \frac{\bm{R}}{R}$ along the $i$-axis ($i=x,y,z$). Note that the Green's tensor is symmetric under $\hat{\bm{R}} \rightarrow - \hat{\bm{R}}$.

\section{\label{app:heatmaps_CDOS} Heatmaps of the CDOS for MD-BIC}

\begin{figure}[h!]
       \centering
        \includegraphics[width=0.8\linewidth]{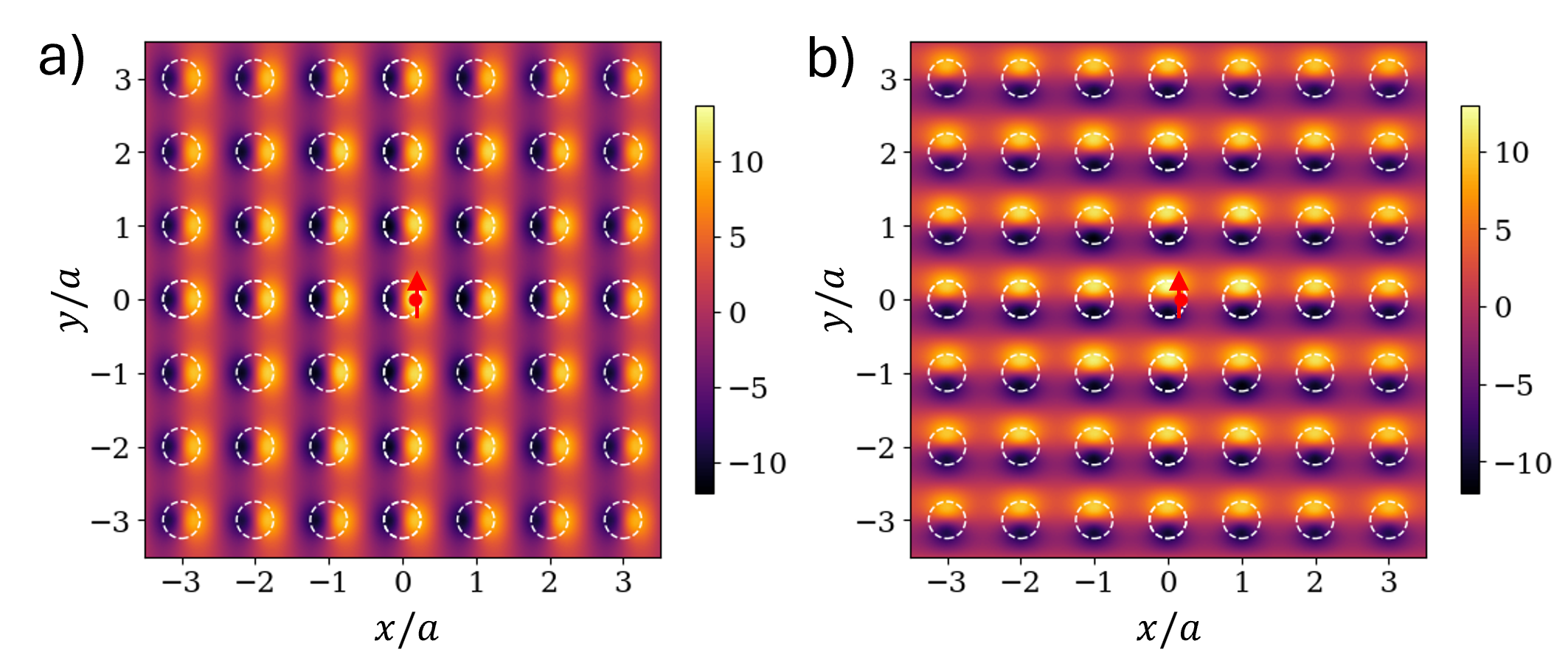} 
    \caption{(a) Collective decay rate $\Gamma_{12}$ (normalized by $\Gamma_0$) with metasurface for the MD-BIC for the same configuration as in Fig.~3(d) in the main text. (b) Same as (a) but now the second emitter has its dipole moment along $x$-axis.}
    \label{fig:supp_info_coll_decay_rate_free_space}
\end{figure}

\section{\label{app:derivation_coll_rates} Derivation of the collective decay rate and $\beta$-factor [Eq.~(3) in the main text]}

\subsection{Approximations over the Green function integral}

The Green function of a system informs us about how the field propagates from one position to another in space. Once the Green function is known, the field scattered by a point dipole source located at $\r_\mu$ with dipole moment $\bm p$ is given by
\begin{equation}
    \bm E_s(\r) = \GG(\r, \r_\mu) \bm p.
\end{equation}
where $\r$ is the observation point.
In the framework of the coupled electric and magnetic dipole (CEMD) \cite{Abujetas20}, the Green function of an \emph{infinite} square periodic metasurface of lattice constant $a$ can be expressed as an integral over the Brillouin zone:
\begin{equation}
    \GG(\r, \r_\mu) = \frac{a^2}{4\pi^2} \int_{\text{1BZ}} \mathrm{d} \kp e^{\imath \kp (\R - \R_\mu)} \GG_l(\kp,\Ro ) \left[1/\alpb - \Gb(\kp)\right]^{-1} \GG_l(\kp, -\Ro_\mu ),
    \label{Eq:1}
\end{equation}
where the dependence on the frequency ($k = \omega/c$) is understood. In this expression, $\r = \R + \Ro$ is the observational point and $\r_\mu = \R_\mu + \Ro_\mu$ the position of the dipole source, where the position is splited in two components: $\R$ is the lattice vector corresponding to the unit cell in which $\r$ is located, while $\Ro$ is the position within the unit cell (and similar for $\R_\mu$ and $\Ro_\mu$) \cite{Zundel2022}, while $\kp$ is the perpendicular component of the wavevector, that is, $\kp = (k_x,k_y)$
(the metasurface is in the $xy$-plane). 
In addition, $\alpb$ is the particle polarizability tensor, $\GG_l(\kp,\r )$ is the lattice Green function (how the field propagates from the metasurface at a given $\kp$) and $\Gb(\kp )$ is the depolarization Green function (how the lattice is depolarized) \cite{Abujetas20}. 

In general, the polarizability and the Green functions are matrices which expressions are relatively complicated. However, we know that, at the BIC frequency, the determinant of the renormalized polarizability $\left[1/\alpb - \Gb(\kp)\right]$ has a real pole at $\kp = (0,0)$. More specifically, the pole is at the $zz$ component of the matrices (electric for MD-BIC and magnetic for ED-BIC), where the sub-index $zz$ refers to the $zz$ component of the matrices. Hence, we can consider that, close to the BIC frequency, the main contribution to the integral of Eq.~(\ref{Eq:1}) is given by the $zz$ component of renormalized polarizability, $\left[1/\alpha_z - G_{b,zz}(\kp)\right]$, around $\kp \sim \bm 0$. In the previous expression, $\alpha_z$ represents the electric or magnetic polarizability along the $z$-axis for MD-BIC and ED-BIC, respectively. 

The ED-BIC state is characterized by an in-phase oscillation of electric dipoles along the $z$-axis.
Therefore, at the ED-BIC frequency, the $z$ component of the electric field scattered by the metasurface, $E_{z}^{\text{ED}}(\r)$, when it is exited by a electric dipole placed at $\r_\mu$ and polarized along the $z$-axis with dipole moment $p$, can be approximated by the following integral
\begin{equation}
    E_{z}^{\text{ED}}(\r) \simeq \frac{a^2}{4\pi^2} \int_{\text{1BZ}} \mathrm{d} \kp e^{\imath \kp (\R - \R_\mu)}G_{l,zz}(\kp,\Ro ) \left[1/\alpha_z - G_{b,zz}(\kp)\right]^{-1} G_{l,zz}(\kp, -\Ro_\mu )p,
    \label{Eq:Ez_BIC}
\end{equation}

Similarly, the MD-BIC state is characterized by an in-phase oscillation of magnetic dipoles along the $z$-axis. Thus, at the MD-BIC frequency, the $y$ component of the electric field scattered by the metasurface, $E_{y}^{\text{MD}}(\r)$, when it is exited by an electric dipole placed at $\r_\mu$ and polarized along the $y$-axis with dipole moment $p$ can be approximated by
\begin{equation}
    E_{y}^{\text{MD}}(\r) \simeq \frac{a^2}{4\pi^2} \int_{\text{1BZ}} \mathrm{d} \kp e^{\imath \kp (\R - \R_\mu)}G_{l,yz}^{(EM)}(\kp,\Ro ) \left[1/\alpha_z - G_{b,zz}(\kp)\right]^{-1} G_{l,yz}^{(EM)}(\kp, -\Ro_\mu )p,
    \label{Eq:Ey_BIC}
\end{equation}
where $G_{l,yz}^{(EM)}(\kp,\r )$ is the component that relates the $y$ component of the magnetic field produced by an electric dipole polarized along the $z$-axis, and vice-versa ($z$ component of the electric field produced by an magnetic dipole polarized along the $y$-axis).

\subsection{Green function expressions}

For completeness, the $zz$ component of the lattice Green functions are \cite{Abujetas20}:
\begin{align}
    G_{l,zz}(\kp,\r ) &= \sum_{n,m=-\infty}^\infty G_{zz}(\bm r, \r_{nm})e^{\imath k_xx}e^{\imath k_yy}\nonumber \\ 
    &= \sum_{l,p=-\infty}^\infty \frac{\imath}{2ab k_z^{(l,p)}} \left[1 - \left(\frac{k_z^{(l,p)}}{k}\right)^2 \right] e^{\imath \left(k_x - \frac{2\pi}{a}l \right)x}e^{\imath \left(k_y - \frac{2\pi}{a}p \right)y}e^{\imath k_z^{(l,p)}|z|},
\end{align}
\begin{align}
    G_{l,yz}^{(EM)}(\kp,\r ) &= \sum_{n,m=-\infty}^\infty G_{yz}^{(EM)}(\bm r, \r_{nm})e^{\imath k_xx}e^{\imath k_yy}\nonumber \\ 
    &= \sum_{l,p=-\infty}^\infty \frac{\imath}{2a^2 k_z^{(l,p)}} \left[- \frac{k_x - \frac{2\pi}{a}l}{k} \right] e^{\imath \left(k_x - \frac{2\pi}{a}l \right)x}e^{\imath \left(k_y - \frac{2\pi}{a}p \right)y}e^{\imath k_z^{(l,p)}|z|},
\end{align}
where $\r_{nm} = (na, ma)$ are the positions of the particles in the metasurface and $k_z^{(l,p)}$ is given by:
\begin{equation}
    k_z^{(l,p)} = \sqrt{k^2 - \left( k_x - \frac{2\pi}{a}l\right)^2 - \left( k_y - \frac{2\pi}{a}p\right)^2}.
\end{equation}

The $zz$ component of the renormalized polarizability reads:
\begin{equation}
    G_{b,zz}(\kp) = \lim_{\r \rightarrow \bm 0} \left[G_{l,zz}(\kp,\r) - G_{zz}(\r,\bm 0) \right].
\end{equation}
In these expressions, $G_{zz}(\r, \r')$ is the $zz$ component of the vacuum Green function:
\begin{equation}
    G_{zz}(\r, \r') = \left(1 + \frac{\partial^2}{\partial z^2} \right)g(\r, \r'), \quad \quad g(\r, \r') = \frac{e^{\imath k |\r - \r'|}}{4\pi|\r - \r'|}.
\end{equation}
and
$G_{yz}^{(EM)}(\r, \r')$ is the $yz$ component of the vacuum Green function
\begin{equation}
    G_{yz}^{(EM)}(\r, \r') = - \frac{\partial}{\partial x}g(\r, \r').
\end{equation}

\subsection{Approximations over renormalized polarizability}

Below diffraction, and in absence of absorption, the imaginary part of the renormalized polarizability appearing in Eqs.~(\ref{Eq:Ez_BIC}) and (\ref{Eq:Ey_BIC}) reads:
\begin{align}
    \Im\left[1/\alpha_z - G_{b,zz}(\kp)\right] = \frac{1}{2a^2 k_z} \left(1 - \frac{k_z^2}{k^2} \right)  \simeq \frac{1}{2a^2 k} \frac{k_\parallel^2}{k^2} ,
\end{align}
where $k_z$ and $k_\parallel$ defined as $k_z \equiv k_z^{(0,0)}$ and $k_\parallel \equiv |\kp|$, and in the last step, we take $k_z \sim k$, where this approximation is valid when $\kp \sim \bm 0$.

On the other hand, the real part does not have a close-form expression, but around the BIC frequency and for $\kp \sim \bm 0$, it can be approximated to 
\begin{align}
     \Re\left[1/\alpha_z - G_{b,zz}(\kp)\right] \simeq A(k_{BIC} - k) -  B\frac{1}{2a^2 k} \frac{k_\parallel^2}{k^2},
\end{align}
where $A$ and $B$ are constants that depend on the specific case. 

Thus, in our range of interest, the depolarization Green function can be approximated as follows:
\begin{equation}
    \left[1/\alpha_z - G_{b,zz}(\kp)\right]^{-1} = \frac{A(k_{BIC} - k) -  B\frac{1}{2a^2 k} \frac{k_\parallel^2}{k^2} + \frac{\imath}{2a^2 k} \frac{k_\parallel^2}{k^2}}{\left[A(k_{BIC} - k) -  B\frac{1}{2a^2 k} \frac{k_\parallel^2}{k^2} \right]^2 + \left[\frac{1}{2a^2 k} \frac{k_\parallel^2}{k^2} \right]^2}.
\end{equation}
Therefore, close to the BIC frequency, the depolarization Green function exhibits a narrow resonance.

\subsection{Approximations over the lattice Green function}

The $zz$ component of the lattice Green function appearing in Eq.~(\ref{Eq:Ez_BIC}) is:
\begin{align}
    G_{l,zz}(\kp,\bm \r ) &= e^{\imath k_x x}e^{\imath k_y y}\sum_{l,p=-\infty}^\infty \frac{\imath}{2ab k_z^{(l,p)}} \left[1 - \left(\frac{k_z^{(l,p)}}{k}\right)^2 \right] e^{-\imath \frac{2\pi}{a}l x}e^{-\imath \frac{2\pi}{a}p y}e^{\imath k_z^{(l,p)}|z|}, \nonumber \\
    & = e^{\imath k_x x}e^{\imath k_y y} \widetilde{G}_{l,zz}(\kp, \r ), \nonumber \\
    & = e^{\imath k_x x}e^{\imath k_y y} \widetilde{G}_{l,zz}(\kp,\bm \rho ),
\end{align}
where for convenience we factor out $e^{\imath k_x x}e^{\imath k_y y}$. In this way, $ \widetilde{G}_{l,zz}(\kp,\bm \rho )$ is periodic, and for this reason we have changed $\r$ by $\bm \rho$ in his evaluation.

Now, let us focus in the properties of $ \widetilde{G}_{l,zz}(\kp,\bm \rho )$ and $\widetilde{G}_{l,zz}(\kp, -\bm \rho_\mu )$, specifically when $\bm\rho = (x,0,z)$ and $\bm\rho_\mu = (0,0,z)$, with $z$ fixed and taking $x$ as the variable (the source is fixed in space and the field is measured along the $x$-axis). First, at $\bm \rho_\mu = (0,0,z)$, the lattice Green function reduce to
\begin{equation}
    \widetilde{G}_{l,zz}(\kp,(0,0,z) ) = \sum_{l,p=-\infty}^\infty \frac{\imath}{2ab k_z^{(l,p)}} \left[1 - \left(\frac{k_z^{(l,p)}}{k}\right)^2 \right] e^{\imath k_z^{(l,p)}|z|}.
\end{equation}
Since $k_z^{(l,p)}$ is purely imaginary $\forall \ (l,p)$ --- except for $(l,p) =(0,0)$ --- below diffraction, the term $(l,p) = (0,0)$ is a complex number, while the other are purely real
\begin{equation}
    \widetilde{G}_{l,zz}(\kp,(0,0,z) ) = \frac{\imath}{2abk_z}\frac{k_\parallel^2}{k^2}e^{\imath k_z |z|} + \sum_{l,p \neq 0} \frac{1}{2ab \widetilde{k}_z^{(l,p)}} \left[1 + \left(\frac{\widetilde{k}_z^{(l,p)}}{k}\right)^2 \right] e^{- \widetilde{k}_z^{(l,p)}|z|},
\end{equation}
with $\widetilde{k}_z^{(l,p)}$ defined as $\widetilde{k}_z^{(l,p)} \equiv -\imath k_z^{(l,p)}$. 

We can see that the complex term is proportional to $k_\parallel^2$. Thus, at small $k_\parallel$ (and for $z$ not too far from the metasurface), the sum term will dominate (and the imaginary part would be much smaller than the real one). \\

N.B.: Notice that, due to the factor $e^{- \widetilde{k}_z^{(l,p)}|z|}$, the dependence on $z$ is exponential. Then, as we go further from the metasurface, the sum would eventually go to zero. This is why in Fig.~2(c) and (d) in the main text, the dependence of the total decay rate (the LDOS) of the first emitter located at position $\r_1=(0,0,z)$, which is proportional to $\Gamma_{11}\propto \Im[E^{\text{ED}}_z(\r_1)/p]$ and hence to $\Re\left[ \widetilde{G}_{l,zz}(\kp,\bm \rho_1 )  \widetilde{G}_{l,zz}(\kp, -\bm \rho_1 )\right]$, was well fitted by an exponential. \newline

 Also, considering that $\frac{2\pi}{a} > k > k_\parallel$, we can assume that $\widetilde{k}_z^{(l,p)}$ does not depend on $k$ nor on $k_\parallel$. With these assumptions, for small $k_\parallel$ and moderate distance $z$, the lattice Green function can be approximated to a real function of $z$
\begin{align}
     \widetilde{G}_{l,zz}(\kp,(0,0,z) ) \sim f(z).
\end{align}
with a small dependence on $k$ and a negligible dependence on $\kp$. \newline

Now, let's see what is the dependence on $x$, by looking at $\bm\rho = (x,0,z)$. In this case, the lattice Green function is
\begin{equation}
    \widetilde{G}_{l,zz}(\kp,(x,0,z)) ) = \sum_{l,p=-\infty}^\infty \frac{\imath}{2ab k_z^{(l,p)}} \left[1 + \left(\frac{k_z^{(l,p)}}{k}\right)^2 \right] e^{-\imath \frac{2\pi}{a}l x} e^{\imath k_z^{(l,p)}|z|},
\end{equation}
and, as before, below diffraction we have:
\begin{equation}
    \widetilde{G}_{l,zz}(\kp,\bm \rho ) = \frac{\imath}{2abk_z}\frac{k_\parallel^2}{k^2}e^{\imath k_z |z|} + \sum_{l,p \neq 0} \frac{1}{2ab \widetilde{k}_z^{(l,p)}} \left[1 + \left(\frac{\widetilde{k}_z^{(l,p)}}{k}\right)^2 \right] e^{-\imath \frac{2\pi}{a}l x} e^{- \widetilde{k}_z^{(l,p)}|z|}.
\end{equation}
Again, the complex propagating term is proportional to $k_\parallel^2$, so we will neglect it. 

Due to that, at $\kp = (0,0)$, $\widetilde{k}_z^{(l,p)}$ is an even function, and as a first approximation we can say that it is also even for small $k_\parallel$. Thus, we get
\begin{align}
    \widetilde{G}_{l,zz}(\kp,\bm \rho ) & \simeq \sum_{l,p = 1}^{\infty} \frac{2}{ab \widetilde{k}_z^{(l,p)}} \left[1 + \left(\frac{\widetilde{k}_z^{(l,p)}}{k}\right)^2 \right] \cos\left(\frac{2\pi}{a}l x\right) e^{- \widetilde{k}_z^{(l,p)}|z|} \nonumber \\
    & + \sum_{l = 1}^{\infty} \frac{1}{ab \widetilde{k}_z^{(l,0)}} \left[1 + \left(\frac{\widetilde{k}_z^{(l,0)}}{k}\right)^2 \right] \cos\left(\frac{2\pi}{a}l x\right) e^{- \widetilde{k}_z^{(l,0)}|z|} \nonumber \\
    & + \sum_{p = 1}^{\infty} \frac{1}{ab \widetilde{k}_z^{(0,p)}} \left[1 + \left(\frac{\widetilde{k}_z^{(0,p)}}{k}\right)^2 \right] e^{- \widetilde{k}_z^{(0,p)}|z|}.
\end{align}
where the propagating contribution is already neglected. 

Similar to what we said before for the case $\bm\rho_\mu = (0,0,z)$, assuming that $\widetilde{k}_z^{(l,p)}$ mainly depend on $l$ and $p$, we can express the lattice Green function as:
\begin{equation}
    \widetilde{G}_{l,zz}(\kp,\bm \rho ) \simeq g(z,0) + \sum_{l = 1}^{\infty} \left( g_0(z,l) + 2g(z,l)\right) \cos\left(\frac{2\pi}{a}l x\right),
\end{equation}
with
\begin{equation}
    g(z,l) = \sum_{p=1}^\infty  \frac{1}{ab \widetilde{k}_z^{'(l,p)}} \left[1 + \left(\frac{\widetilde{k}_z^{'(l,p)}}{k}\right)^2 \right] e^{- \widetilde{k}_z^{'(l,p)}|z|},
\end{equation}
\begin{equation}
    g_0(z,l) = \frac{1}{ab \widetilde{k}_z^{'(l,0)}} \left[1 + \left(\frac{\widetilde{k}_z^{'(l,0)}}{k}\right)^2 \right] e^{- \widetilde{k}_z^{'(l,0)}|z|},
\end{equation}
where the prime in $\widetilde{k}_z^{'(l,p)}$ indicates that we take $\kp = (0,0)$, i.e., 
\begin{equation}
    \widetilde{k}_z^{(l,p)} \sim\widetilde{k}_z^{'(l,p)} \equiv \sqrt{\frac{4\pi^2}{a^2}(l^2 + p^2) - k^2}.
\end{equation}
Then, the real function $g(z,l)$ (and $g_0(z,l)$) determines the contribution of each cosine to the sum. \newline

Specifically, for the ED-BIC of silicon Mie-spheres of $n=3.5$, and if we take $y = 0$ and $z = 104$ nm like the configuration in the main text, the cosine expansion of the field can be calculated as:
\begin{equation}
     \widetilde{G}_{l,zz}(\kp,(x,0,z) ) \propto 0.53 + \cos\left(\frac{2\pi}{a} x\right) + 0.31\cos\left(\frac{4\pi}{a} x\right) + 0.0936\cos\left(\frac{6\pi}{a} x\right) + 0.00258\cos\left(\frac{8\pi}{a} x\right) + \cdots .
     \label{eq:cos_exp_EDBIC}
\end{equation}
\newline

Following a similar derivation, for the MD-BIC case, it is possible to write the $yz$ (electric-magnetic) component of the lattice Green function as
\begin{align}
    G_{l,yz}^{(EM)}(\kp,\bm \r ) &= e^{\imath k_x x}e^{\imath k_y y}\sum_{l,p=-\infty}^\infty \frac{\imath}{2ab k_z^{(l,p)}} \left[- \frac{k_x - \frac{2\pi}{a}l}{k} \right] e^{\imath k_x - \frac{2\pi}{a}l x}e^{\imath - \frac{2\pi}{a}p y}e^{\imath k_z^{(l,p)}|z|}, \nonumber \\
    & = e^{\imath k_x x}e^{\imath k_y y} \widetilde{G}_{l,yz}^{(EM)}(\kp,\bm \rho ),
\end{align}
Below diffraction, we have
\begin{equation}
    \widetilde{G}_{l,yz}^{(EM)}(\kp,\bm \rho ) = -\frac{\imath}{2abk_z}\frac{k_x}{k}e^{\imath k_z|z|} + \sum_{l,p \neq 0} \frac{1}{2ab \widetilde{k}_z^{(l,p)}} \left[- \frac{k_x - \frac{2\pi}{a}l}{k} \right] e^{-\imath \frac{2\pi}{a}l x}e^{-\imath \frac{2\pi}{a}p y} e^{- \widetilde{k}_z^{(l,p)}|z|},
\end{equation}
and assuming that at $\kp = (0,0)$, $\widetilde{k}_z^{(l,p)}$ is a even function, as a first approximation we can say that 
\begin{align}
    \widetilde{G}_{l,yz}^{(EM)}(\kp,\bm \rho ) \approx -\frac{\imath}{2abk_z}\frac{k_x}{k}e^{\imath k_z|z|} &+ 2\imath \sum_{l,p \neq 0} \frac{1}{ab \widetilde{k}_z^{(l,p)}} \left[- \frac{\frac{2\pi}{a}l}{k} \right] \sin\left(\frac{2\pi}{a}l x\right)\cos\left( \frac{2\pi}{a}p y\right) e^{- \widetilde{k}_z^{(l,p)}|z|} \nonumber \\
    &- 2\sum_{l,p \neq 0} \frac{1}{ab \widetilde{k}_z^{(l,p)}} \left[- \frac{k_x}{k} \right] \cos\left(\frac{2\pi}{a}l x\right)\cos\left( \frac{2\pi}{a}p y\right) e^{- \widetilde{k}_z^{(l,p)}|z|},
\end{align}
and further assuming that $k_x$ is small
\begin{equation}
        \widetilde{G}_{l,yz}^{(EM)}(\kp,\bm \rho ) \approx  2\imath \sum_{l,p \neq 0} \frac{1}{ab \widetilde{k}_z^{(l,p)}} \left[- \frac{\frac{2\pi}{a}l}{k} \right] \sin\left(\frac{2\pi}{a}l x\right)\cos\left( \frac{2\pi}{a}p y\right) e^{- \widetilde{k}_z^{(l,p)}|z|}.
        \label{eq:gyz}
\end{equation}
Specifically, for the MD-BIC of silicon Mie-spheres of $n=3.5$, and if we take $\bm \rho=(x,y,z)$ with $x = 0.164a$ and $z = 104$ nm like the configuration in the main text, the cosine expansion of the field can be calculated as:
\begin{equation}
     \widetilde{G}_{l,yz}^{(EM)}(\kp,(0.164a,y,z) ) \propto 1 + 0.547\cos\left(\frac{2\pi}{a} y\right) + 0.0088\cos\left(\frac{4\pi}{a} y\right) + 0.00014\cos\left(\frac{6\pi}{a} y\right) + \cdots .
     \label{eq:cos_exp_MDBIC}
\end{equation}

Note that in both cases, in first order, the lattice Green functions do not depend on $\kp$.

\subsection{Cross density of states integral}

In base of the previous approximations, the cross density of states (CDOS) associated to the ED-BIC is proportional to $\Gamma_{12} \propto \Im[E^{\text{ED}}_z(\r)/p]$ (the metasurface is exited by an electric dipole along the $z$ axis),
\begin{align}
    \Im[E^{\text{ED}}_{z}(\r)/p] \simeq \frac{a^2}{4\pi^2} \int_{\text{1BZ}} \mathrm{d} \kp \Im\left\lbrace e^{\imath k_x x} \left[1/\alpha_z - G_{b,zz}(\kp)\right]^{-1} \right\rbrace \Re\left[ \widetilde{G}_{l,zz}(\kp,\bm \rho )  \widetilde{G}_{l,zz}(\kp, -\bm \rho_\mu ) \right] 
\end{align}
where we have already neglected the imaginary part of $\widetilde{G}_{l,zz}(\kp,\bm \rho )$. 

Since the depolarization Green function represents a narrow resonance around $\kp \sim \bm 0$, and the lattice Green functions do not depend on $\kp$ at small $\kp$, we can separate them as   
\begin{align}
    \Im[E^{ED}_{z}(\r)/\mu] \simeq \frac{a^2}{4\pi^2} \Re\left[ \widetilde{G}_{l,zz}(\kp,\bm \rho )  \widetilde{G}_{l,zz}(\kp, -\bm \rho_\mu ) \right] \int_{\text{1BZ}} \mathrm{d} \kp \Im\left\lbrace e^{\imath k_x x} \left[1/\alpha_z - G_{b,zz}(\kp)\right]^{-1} \right\rbrace . 
\end{align}
and with the help of the next integrals
\begin{equation}
    \int_{0}^{2\pi} \cos\left(k_\parallel (\cos\theta x + \sin\theta y)\right) \mathrm{d} \theta = 2\pi J_0\left(k_\parallel \sqrt{x^2 + y^2}\right),
\end{equation}
\begin{equation}
    \int_{0}^{2\pi} \sin\left(k_\parallel (\cos\theta x + \sin\theta y)\right) \mathrm{d} \theta = 0,
\end{equation}
to show that
\begin{align}
    \Gamma_{12}(x) & \propto \Re\left[ \widetilde{G}_{l,zz}(\kp^{\text{res}},\bm \rho )  \widetilde{G}_{l,zz}(\kp^{\text{res}}, -\bm \rho_\mu ) \right] \times J_0(k_\parallel^{\text{res}}|x|) \nonumber \\
    &\sim \left[\sum \gamma_j \cos\left( j\frac{2\pi}{a}x\right) \right] \times J_0(k_\parallel^{\text{res}}|x|),
    \label{eq:gamma12_EDBIC}
\end{align}
where $\kp^\text{res}$ is the $\kp$ at which the depolarization Green function resonates and $k_\parallel^{\text{res}}\equiv|\kp^\text{res}|$ (closer to zero as the frequency is closer to the BIC frequency), and the coefficients $\gamma_j$, when particularized for the ED-BIC of silicon Mie-spheres, are the ones from Eq.~(\ref{eq:cos_exp_EDBIC}). \newline

Similarly, the CDOS associated to the MD-BIC is proportional to $\Gamma_{12} \propto \Im[E^{\text{MD}}_y(\r)/p]$ (the metasurface is exited by an electric dipole along the $y$ axis), and with a similar procedure as for the ED-BIC, we get 
\begin{align}
    \Gamma_{12}(y) & \propto \Re\left[ \widetilde{G}_{l,yz}^{(EM)}(\kp^{\text{res}},\bm \rho )  \widetilde{G}_{l,yz}^{(EM)}(\kp^{\text{res}}, -\bm \rho_\mu ) \right] \times J_0(k_\parallel^{\text{res}}|y|) \nonumber \\
    &\sim \left[\sum \gamma_j \cos\left( j\frac{2\pi}{a}y\right) \right] \times J_0(k_\parallel^{\text{res}}|y|),
    \label{eq:gamma12_MDBIC}
\end{align}
where the coefficients $\gamma_j$, when particularized for the MD-BIC of silicon Mie-spheres, are the ones from Eq.~(\ref{eq:cos_exp_MDBIC}). Eqs.~(\ref{eq:gamma12_EDBIC}) and (\ref{eq:gamma12_MDBIC}) are defined up to a pre-factor (undetermined), that we will call "$C$" in the following.

\subsection{$\beta$-factor}

The most important figure-of-merit in this paper is the $\beta$-factor. Similar to photonic crystal waveguides \cite{lecamp2007very,manga2007single,arcari2014near}, we define here the single-emitter $\beta$-factor associated to the BIC mode, which gives the probability of the photon to be emitted into the BIC mode. 
The BIC $\beta$-factor is in fact given by Eqs.~(\ref{eq:gamma12_EDBIC}) and (\ref{eq:gamma12_MDBIC}) in the limit when $x= 0$ and $y= 0$ for the ED- and MD-BIC, respectively (since the CDOS corresponds to the LDOS for $0$ separations), and normalized by $\Gamma$, the total single-emitter decay rate:
\begin{equation}
\beta = C\frac{\sum \gamma_j}{\Gamma}
\label{eq:gammaBIC_expression}
\end{equation}
with $C$ a pre-factor.

By next expressing the pre-factor $C$ in terms of the $\beta$-factor using Eq.~(\ref{eq:gammaBIC_expression}), Eq.~(3) in the main text is obtained from Eqs.~(\ref{eq:gamma12_EDBIC}) and (\ref{eq:gamma12_MDBIC}) (note that in Eq.~(3), the coefficients of the cosine expansions have been renormalized by the quantities $\sum \gamma_j$).

\begin{figure}[h!]
       \centering
        \includegraphics[width=0.8\linewidth]{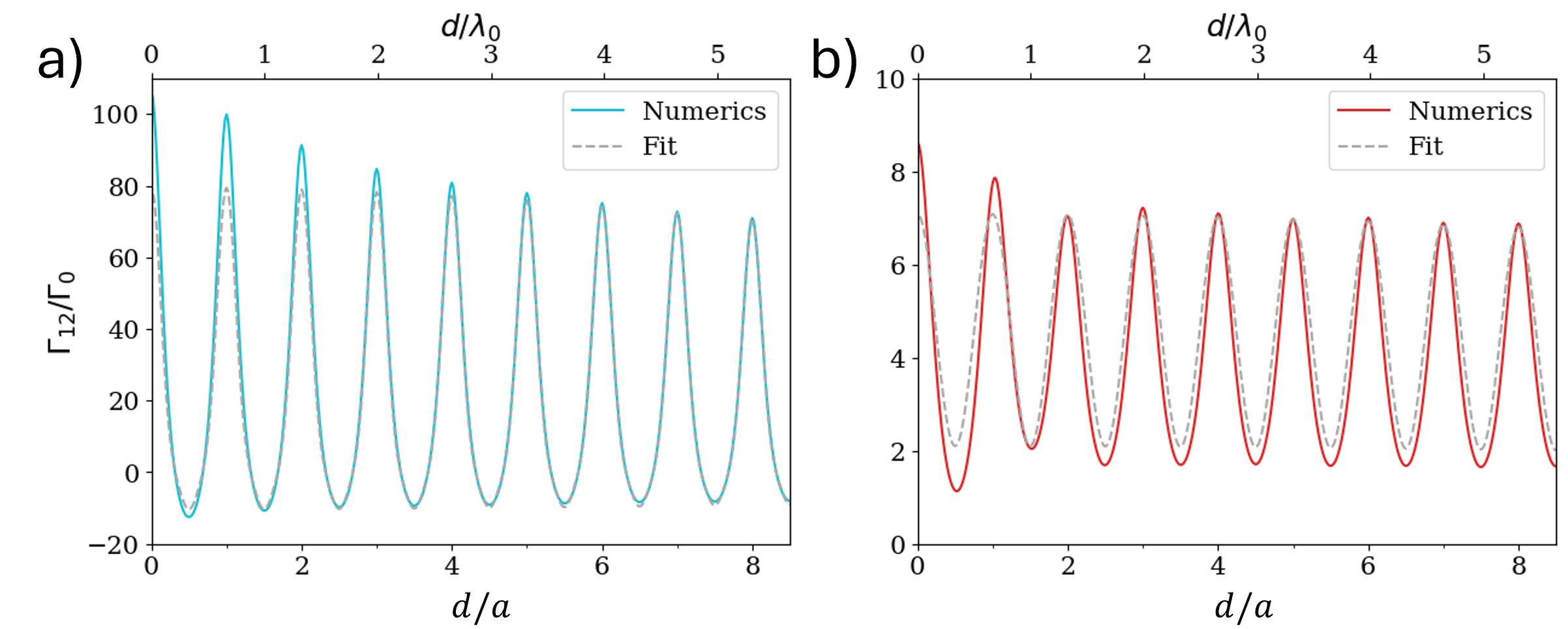} 
    \caption{Same as Figs.~4(a) and 4(b) in the main text but for the \emph{infinite} metasurface.}
    \label{app:fig:supp_info_2}
\end{figure}

We plot these analytical expressions in Fig.~\ref{app:fig:supp_info_2} (dashed grey curves) together with numerical simulations for an infinite metasurface (solid curves) computed using CEMD method (see Section~\ref{app:methods}). The agreements are very good. The fitting parameters are: $\beta = 75.18\%$ and $k_\parallel^{\text{res}}=0.219\,$rad/$\mu$m for the ED-BIC and $\beta = 82.43\%$ and $k_\parallel^{\text{res}}=0.125\,$rad/$\mu$m for the MD-BIC. Non-zero values for $k_\parallel^{\text{res}}$ are used to account for the numerical detuning --- the fact that it is difficult to find exactly the wavelength that matches with the true BIC wavelength due to the very narrow resonance line of ideal BIC --- even if in the infinite case, $k_\parallel^{\text{res}}=0$ for strict wavelength match.

\begin{figure}[h!]
       \centering
        \includegraphics[width=0.8\linewidth]{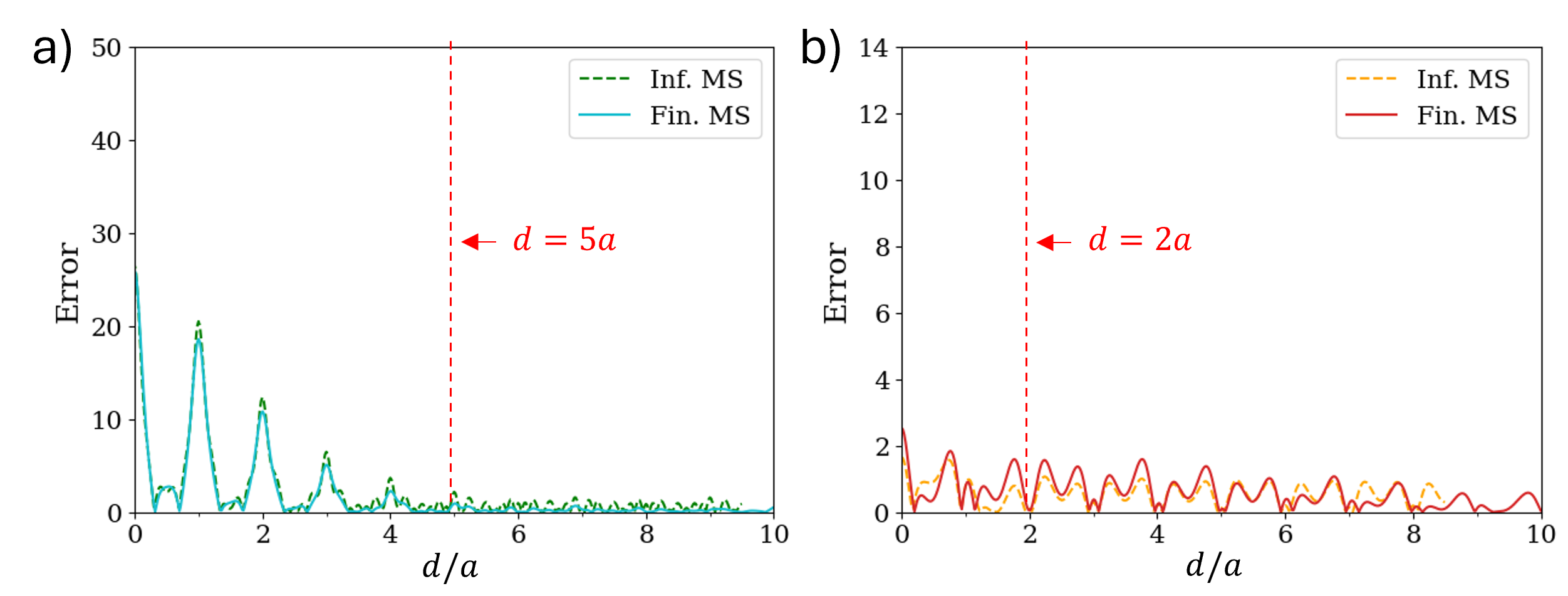} 
    \caption{Absolute error between numerics and fits for the collective decay rate, in the case of (a) the ED-BIC and (b) the MD-BIC. ``Inf. MS" refers to infinite metasurface, and ``Fin. MS" refers to finite metasurface.}
    \label{fig:supp_info_error}
\end{figure}

In Fig.~\ref{fig:supp_info_error}, we plot the absolute error between the numerical and fitted collective decay rates for both the ED-BIC and the MD-BIC, in the infinite case (see Fig.~\ref{app:fig:supp_info_2}) and the finite case (see Fig.~4 of the main text). One can see that the agreement becomes very good for $d\geq 5a$ in the ED-BIC case and $d\geq 2a$ in the MD-BIC case.

\newpage
\section{\label{app:coeff} Plots of coefficients $\Gamma_{22}$ and $\Omega_{12}$}

\begin{figure}[h!]
       \centering
        \includegraphics[width=0.8\linewidth]{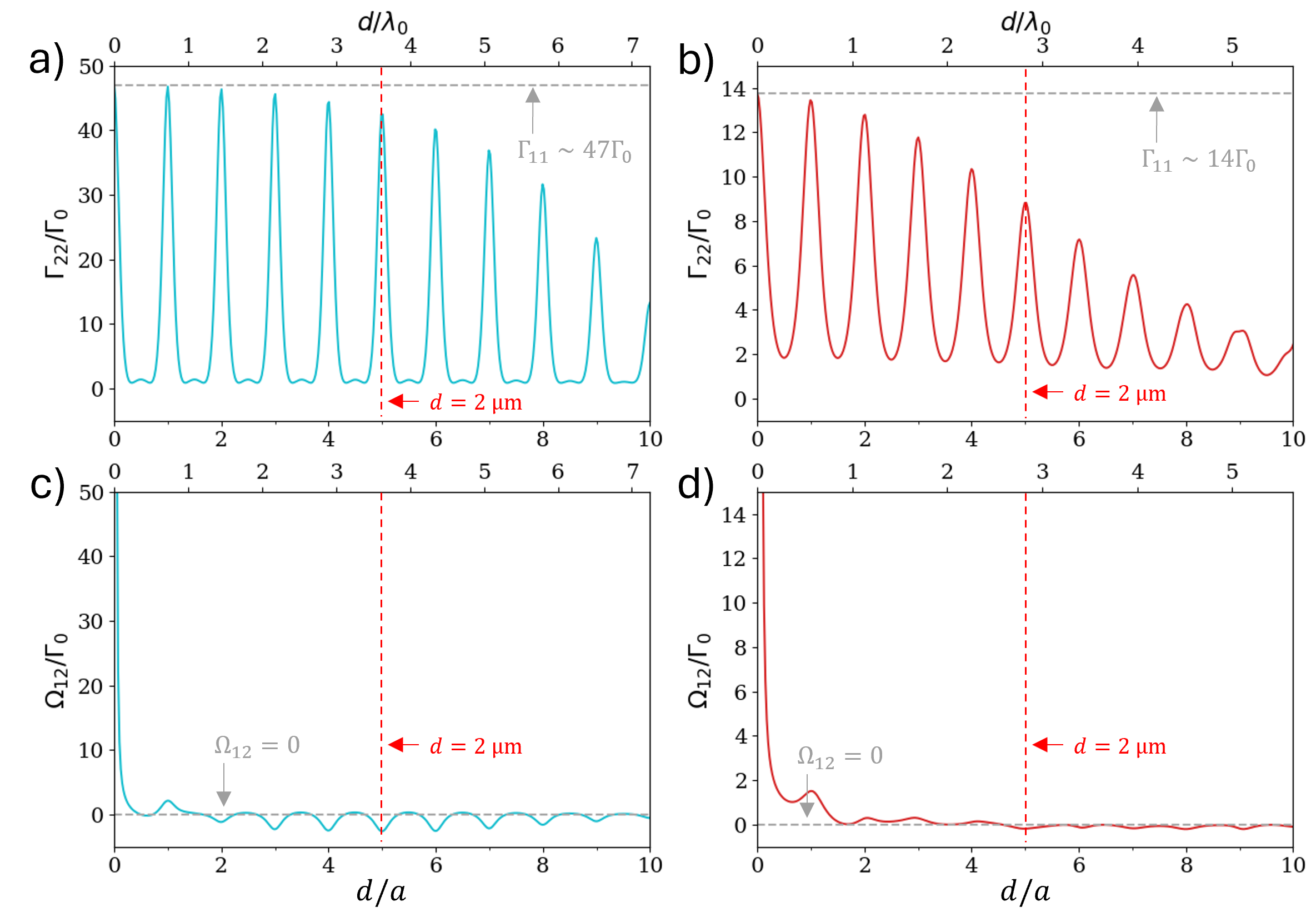} 
    \caption{(a) and (b): Decay rate $\Gamma_{22}$ of the second quantum emitter as a function of emitter separation $d$ for (a) the ED-BIC and (b) the MD-BIC. (c) and (d): Collective coupling rate $\Omega_{12}$ as a function of $d$ for (c) the ED-BIC and (d) the MD-BIC. The rates are normalized by the free space single emitter decay rate $\Gamma_0$, and are plotted along the same line-cuts as in Fig.~3(c) and (d) of the main text.}
    \label{fig:supp_info_decay_rates}
\end{figure}

